\documentclass{emulateapj}
\usepackage{lscape}

\newcommand{\ug}{$u'-g'$ }
\newcommand{\gi}{$g'-i'$ }
\newcommand{\iz}{$i'-z'$ }
\newcommand{\ugiz}{$u'g'i'z'$ }
\newcommand{\av}{$A_V$ }

\begin{document}

\title{THE INITIAL CLUSTER MASS FUNCTION OF SUPER STAR CLUSTERS IN IRREGULAR AND SPIRAL GALAXIES}
\author{Jayce D. Dowell$^1$, Brent A. Buckalew$^{2,4}$, and Jonathan C. Tan$^{3,5}$}
\affil{$^1$Department of Astronomy, Indiana University, 727 E. 3$^{rd}$ St, Bloomington, IN  47405, USA}
\affil{$^2$Department of Physics, Embry-Riddle Aeronautical University, 3700 Willow Creek Rd, Prescott, AZ  86301, USA}
\affil{$^3$Institute of Astronomy, Department of Physics, ETH Z\"urich, 8093 Z\"urich, Switzerland}
\email{jdowell@astro.indiana.edu}
\email{buckaleb@erau.edu}
\email{jt@astro.ufl.edu}

\altaffiltext{4}{Previous Address:\\Dept. of Physics \& Astronomy\\University of Wyoming\\1000 E. University Ave.\\Laramie, WY 82071}
\altaffiltext{5}{Current Address:\\ Dept. of Astronomy\\University of Florida\\Gainesville, FL 32611}

\journalinfo{The Astronomical Journal \textbf{135} (2008) 823}
\received{2007 May 26}
\accepted{2007 October 31}
\submitted{Published 2008 February 5}

\begin{abstract}
The initial cluster mass function (ICMF) is a fundamental property of
star formation in galaxies. To gauge its universality, we measure and
compare the ICMFs in irregular and spiral galaxies. Our sample of
irregular galaxies is based on 13 nearby galaxies selected from
a volume-limited sample from the fifth data release of the Sloan
Digital Sky Survey (SDSS), from which about 320 young ($\leq 20$~Myr),
massive ($\gtrsim 3\times 10^4\:M_\odot$) clusters or associations were
selected using an automated source extraction routine. The
extinctions, ages, and masses were determined by comparing their \ugiz
magnitudes to those generated from starburst models. Completeness
corrections were performed using Monte Carlo simulations in which
artificial clusters were inserted into each galaxy. Foreground stellar
and background galactic contaminations were assessed by analyzing SDSS
images of fields around the sample galaxies and found to be small. We
analyzed three nearby spiral galaxies with SDSS data exactly in the
same way to derive their ICMF based on a similar number of young,
massive clusters as the irregular galaxy ICMF. We find that the ICMFs
of irregular and spiral galaxies for masses $>10^{4.4}\:M_\odot$ are
statistically indistinguishable.  For clusters and associations more
massive than $10^{4.4}\:M_\odot$, the ICMF of the irregular galaxies is
reasonably well fit by a power law $\frac{dN(M)}{dM} \varpropto
M^{-\alpha_M}$ with $\alpha_M = 1.88 \pm 0.09$. Similar results were
obtained for the ICMF of the spiral galaxy sample but with $\alpha_M =
1.75 \pm 0.06$. We discuss the implications of our results on theories
of star cluster formation, which appear to indicate that the power-law 
indicies are independent of metallicity and galactic shear rate.  We also 
examine the evolution of visual extinction, $A_V$, with cluster age and 
find significant reduction in median extinction after $\sim 5-10$~Myr 
by about 0.5~mag for clusters in both spiral and irregular galaxies.  We 
discuss the implications of our results for theories of star cluster 
formation: the shape of the ICMF appears to be independent of metallicity 
and galactic shear rate.
\end{abstract}


\section{Introduction}

Star formation is a highly clustered process. A large fraction of
stars form in clusters, as determined in our local Galactic
neighborhood \citep{LL03}, in interacting galaxies like the
Antennae \citep{Fall05}, and in dwarf starburst
galaxies such as NGC~5253 \citep{Tremonti01}. The most massive
star clusters, $\gtrsim 10^5\:M_\sun$, are known as super star
clusters (SSCs), and examples are found in all of these galactic types
\citep{CNCG05, MLTG02, Turner00}.

The study of SSCs is important for a number of reasons. Their extreme
physical properties present new challenges for star-formation
theories.  For example, their large masses and relatively small sizes
mean that their natal gas clouds must have very high pressures due to
self-gravity and escape speeds that are in excess of the ionized gas
sound speed \citep{tan05}. The stellar densities of the final clusters
are not likely to be that much greater than typical rich young
Galactic star clusters, such as the Orion Nebula Cluster, but if the
initial mass function (IMF) is similar, then there are more massive OB
stars present, potentially thousands within just a few parsecs of each
other.

Young SSCs are likely to be similar to the progenitors of globular
clusters \citep{GD97, ZF99} although this requires an evolution of
the initial cluster mass function (ICMF) from a power-law form to a
log-normal form, likely to occur through preferential disruption of
lower-mass clusters. Thus, studying nearby populations of SSCs may shed
light on processes similar to those occurring in the early phases of
galaxy formation.

From a practical point of view, young SSCs are the easiest clusters to
identify in magnitude limited surveys of external galaxies, and so
corrections for incompleteness are relatively small, allowing for more
accurate determination of the ICMF. However,
a disadvantage is that SSCs are relatively rare, so a large sample of
galaxies needs to be surveyed to build up a good statistical sample.

While SSCs occur in many different types of galactic environments,
there are many prominent examples in irregular and dwarf irregular galaxies. For
example, NGC~5253 hosts a $10^6\:M_\odot$ SSC \citep{Turner00},
Henize $2-10$ hosts several forming SSCs as revealed in the radio and mm
continuum \citep{KJ99,JK03}, NGC~1569 hosts three SSCs \citep{GG03}, SBS~0335-052 hosts a
$2\times 10^6\:M_\odot$ cluster \citep{PS02}, and NGC~1705
and IC~4662 host SSCs \citep{JIP03}. Several other dwarf starburst galaxies 
with SSCs are described by \citet{Beck00}.

\citet{BHE02} discussed how the most massive cluster in a galaxy
depends on the mass function and the size of the galaxy \citep[see
also][]{HEDM03}. They argued that the most massive clusters in dwarf
irregular galaxies are more massive than would be expected if their
star cluster populations are similar to those of normal disk galaxies.
They suggested that this result could imply that the ICMF of dwarf
galaxies is not a continuous power law and that the SSCs have a
different origin perhaps related to special galactic circumstances,
such as an interaction with another galaxy. However, care needs to be
taken with the selection of the galaxy sample; for example, if only
dwarf irregular galaxies with prominent SSCs are considered, then the
result will be biased.  The continuity of the ICMF in dwarf galaxies
is an issue that we specifically address in this paper.  It is
important to quantify potential differences in the ICMFs in different
types of galaxies since it may illustrate how the star-formation
process responds to environmental factors.

To characterize properly the masses of a population of star clusters,
we need to measure the ICMF. We shall
see that it can often be well fit by a power law,
	\begin{equation}
	\frac{dN(M)}{dM} \varpropto M^{-\alpha_M}.
	\end{equation}
The masses of clusters are not easily determined
observationally. Dynamical masses from spectroscopic studies have been
possible only for a handful of SSCs \citep{MLTG02,MGG03}.  Most
previous studies of cluster populations have concentrated on the
luminosity function (LF), which also often resembles a power law,
	\begin{equation}
	\frac{dN(L)}{dL} \varpropto L^{-\alpha_L}.
	\end{equation}
From a sample of spiral galaxies, \citet{L02} found
$2.0\lesssim\alpha_L \lesssim 2.4$. In the Antennae galaxies,
\citet{WZLFSM99} found $\alpha_L\simeq 2.1$.  \citet{Fall06} has 
discussed the relations between luminosity and mass functions in 
more detail.

For unresolved clusters, conversion from luminosity to mass requires
knowledge of the age, metallicity, and reddening of the cluster
together with a model for the cluster's stellar evolution and an
assumed IMF for the stars. Multi-color
photometry is a minimum observational requirement for performing this
conversion on a cluster-by-cluster basis and is the approach adopted
in this paper using Sloan Digital Sky Survey (SDSS)
data. \citet{BHE02} performed a similar analysis for nearby dwarf
irregular galaxies using a heterogeneous dataset from the \textit{HST}
archives. Their study has the advantage of high spatial resolution,
but the disadvantage of a limited spectral coverage (and thus more of
an uncertain conversion from luminosity to mass). \citet{HEDM03} used
ground-based data to study the cluster populations of the Large and
Small Magellanic Clouds, concluding that $2.0\lesssim \alpha_M\la
2.4$.  \citet{ZF99} found $\alpha_M \simeq 2$ for young (${\la} 160$
Myr) clusters with masses $10^4 \: \la \: M \: \la \: 10^6\:M_\sun$ in the
merging galaxies NGC~4038/4039 using multi-wavelength, broadband \textit{HST}
photometry. Based on the LFs of Galactic HII regions, \citet{MCK97} 
concluded $\alpha_M \simeq 2$ for clusters and associations.

Alternatively, the conversion from luminosity to mass can be made on a
statistical basis, given assumptions about the star-formation
history. \citet{L02} showed that a cluster population with mass
function with $\alpha_M=2$ from $10^3\:M_\odot< M <10^5\:M_\odot$ and a
uniform distribution of ages between $10^7$ and $10^9\:{\rm yr}$ has an
approximately power-law LF in the range $-11<M_V<-8$ with $\alpha_L=2.72$. 

Comparisons of ICMFs in different galaxy types are interesting because it
informs us about the star-formation process and how it may depend on
different physical conditions. For example, \citet{EE97} argued for a
universal formation mechanism for open and globular clusters because
the ICMF was similar to the mass spectrum of interstellar
clouds. Differences in the mass function between dwarf galaxies and
normal disk galaxies would cast doubt upon this assertion. Dwarf
galaxies have quite different physical properties compared to normal
disk galaxies such as typically lower metallicities and less shear
due to differential rotation, and the influence of these factors on
the star-formation process may imprint itself on the ICMF.

In this paper, we determine the ICMF for 13 irregular galaxies
within 10~Mpc with data obtained from the fifth data release (DR5) of
the SDSS. We compare this to the ICMF derived from SDSS DR5 data of
three spiral galaxies, also within 10~Mpc.  In Section 2, we discuss the SDSS 
DR5 data, the selection of our two galaxy samples (irregulars and 
spirals), and the procedures for finding clusters and deriving their 
magnitudes. Section 3 deals with extinction, age, and mass determinations 
of the individual star clusters.  We consider the ICMF in Section 4; in 
particular we ask if the ICMFs of irregular and spiral galaxies are 
consistent with being drawn from the same parent distribution. We fit 
the observed ICMFs with power laws and derive their parameters.  We 
discuss the implications of our findings in Section 5 and summarize in Section 6.

\section{SDSS Data and Photometry}
\subsection{Sample Selection}

\begin{figure}
	\plottwo{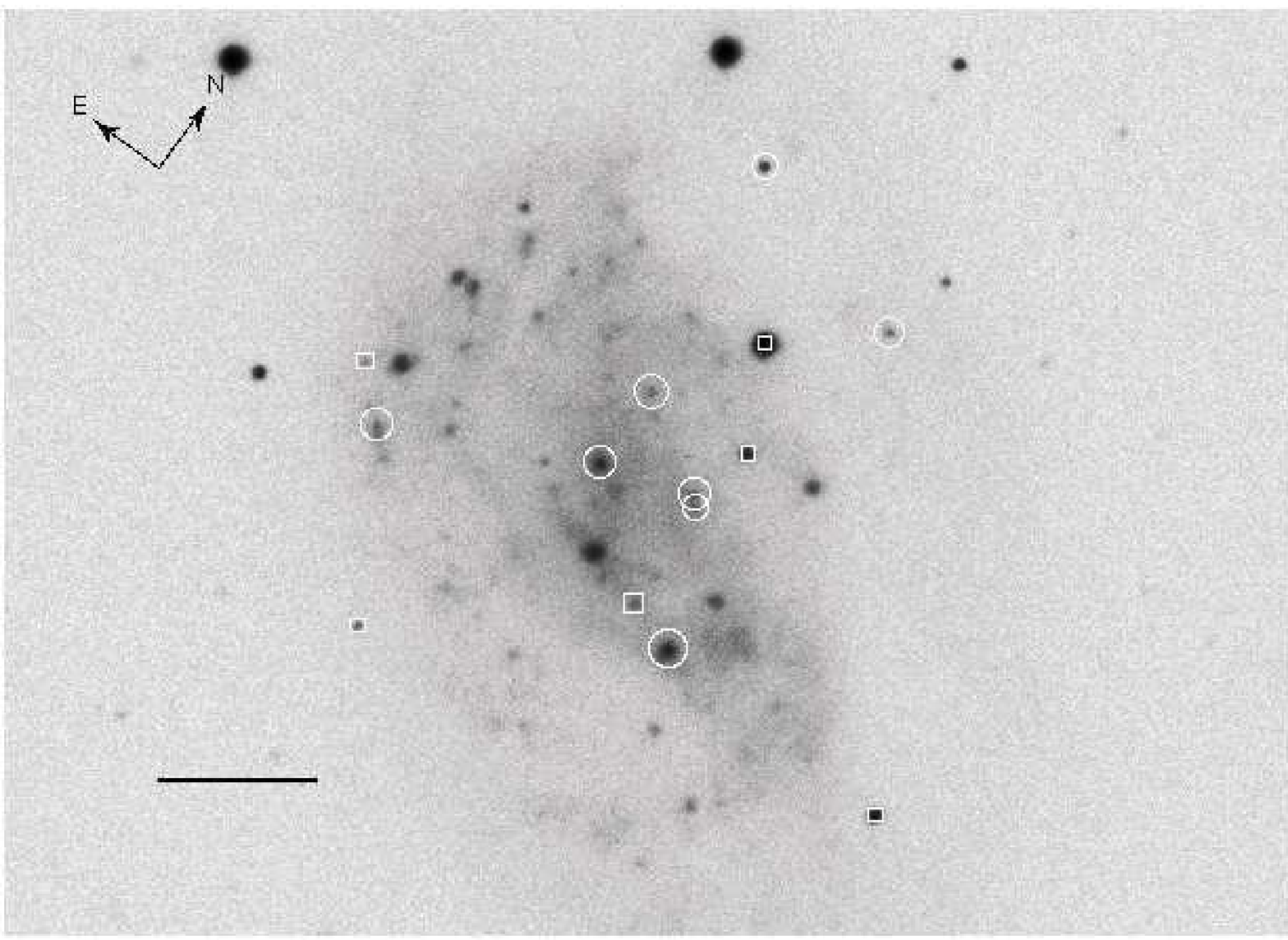}{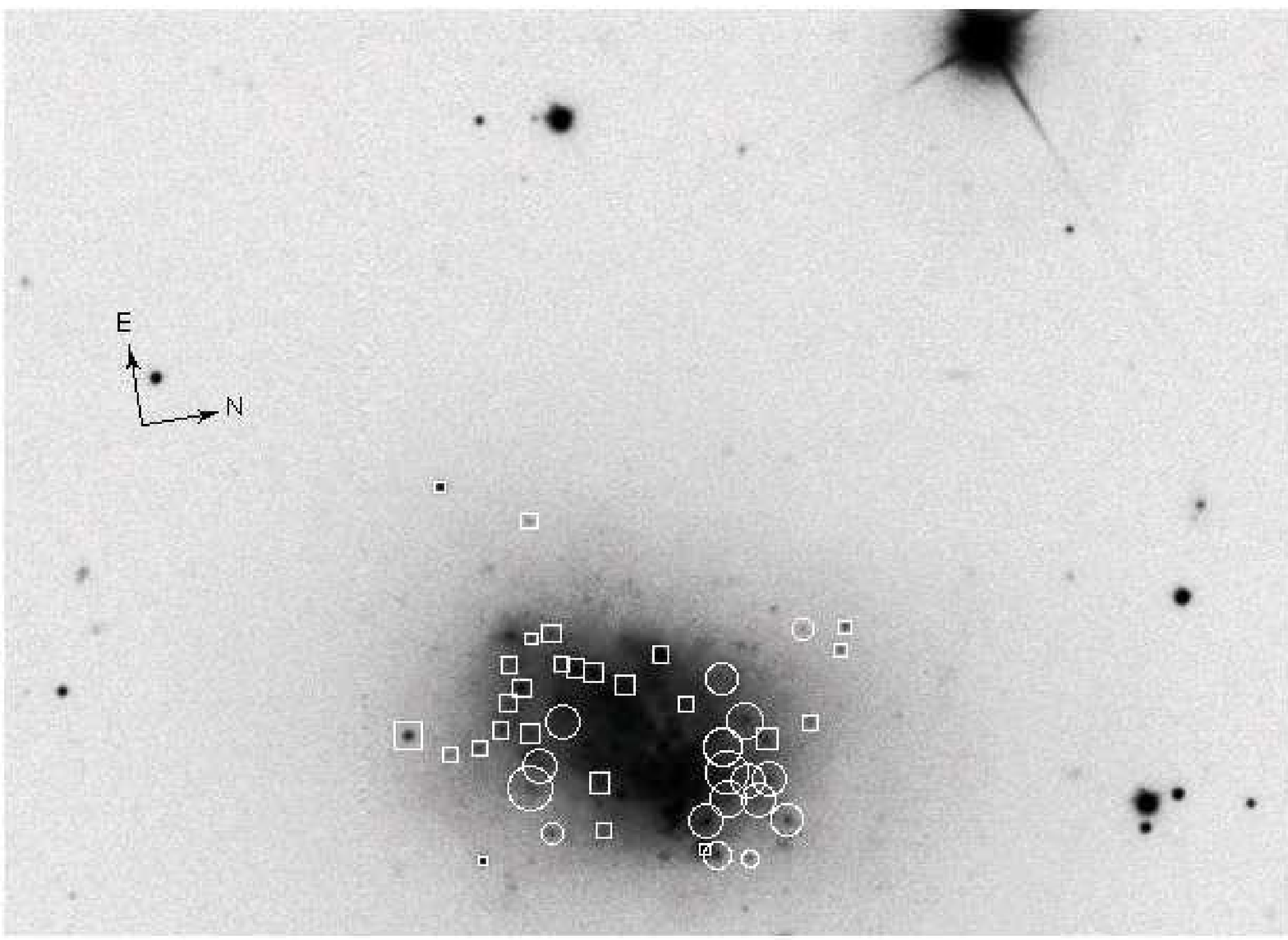}
	\plottwo{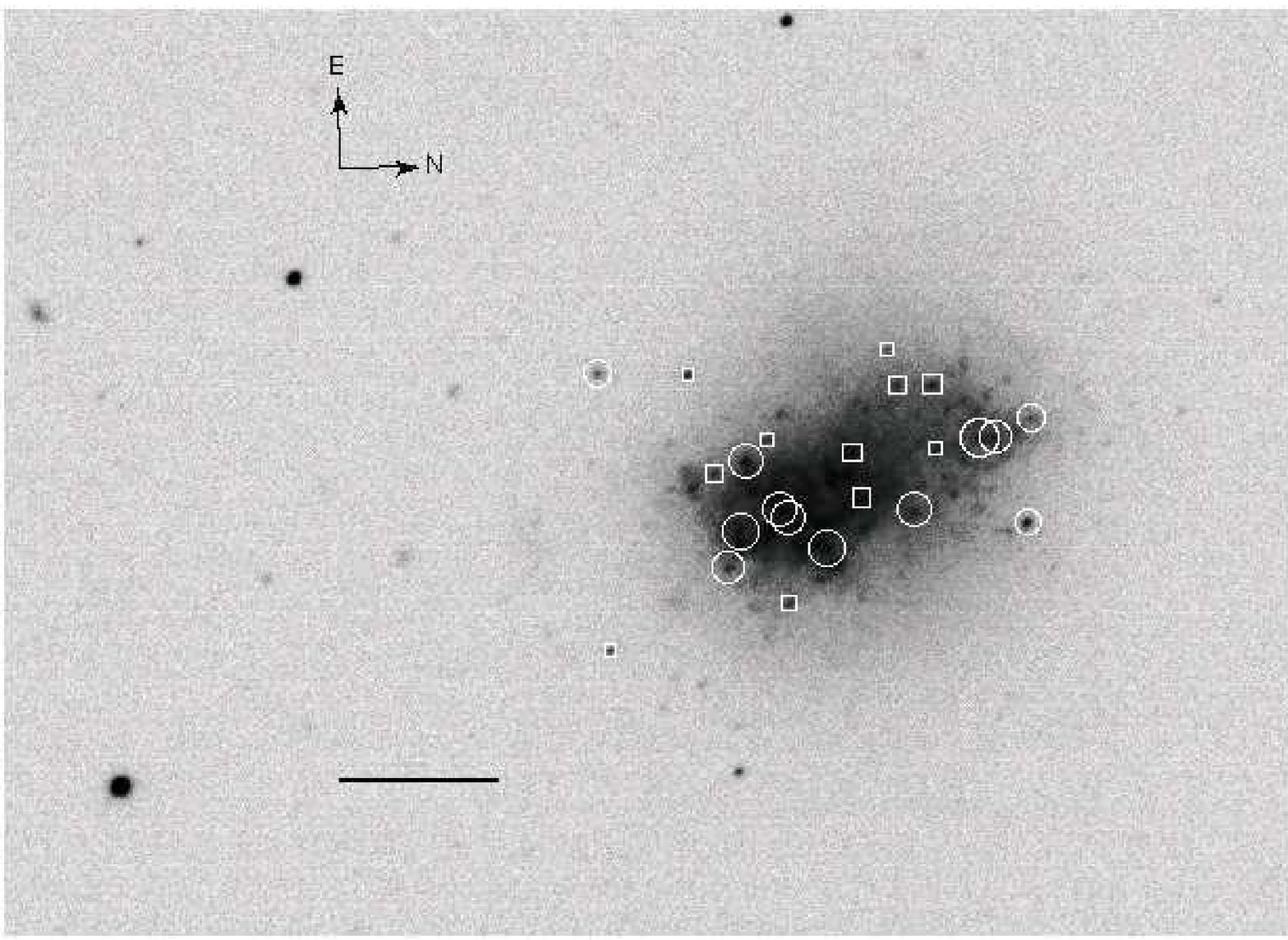}{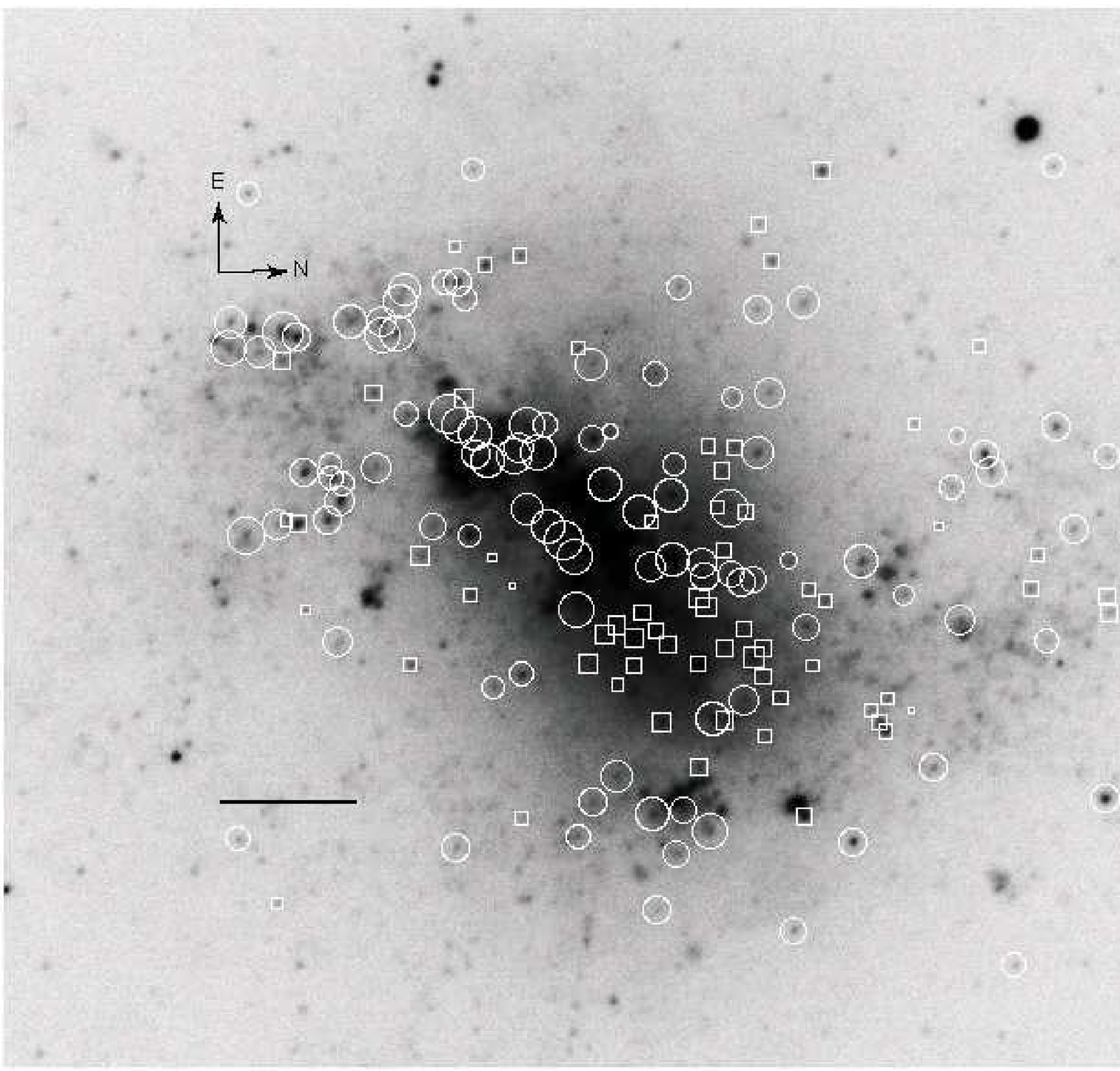}
	\plottwo{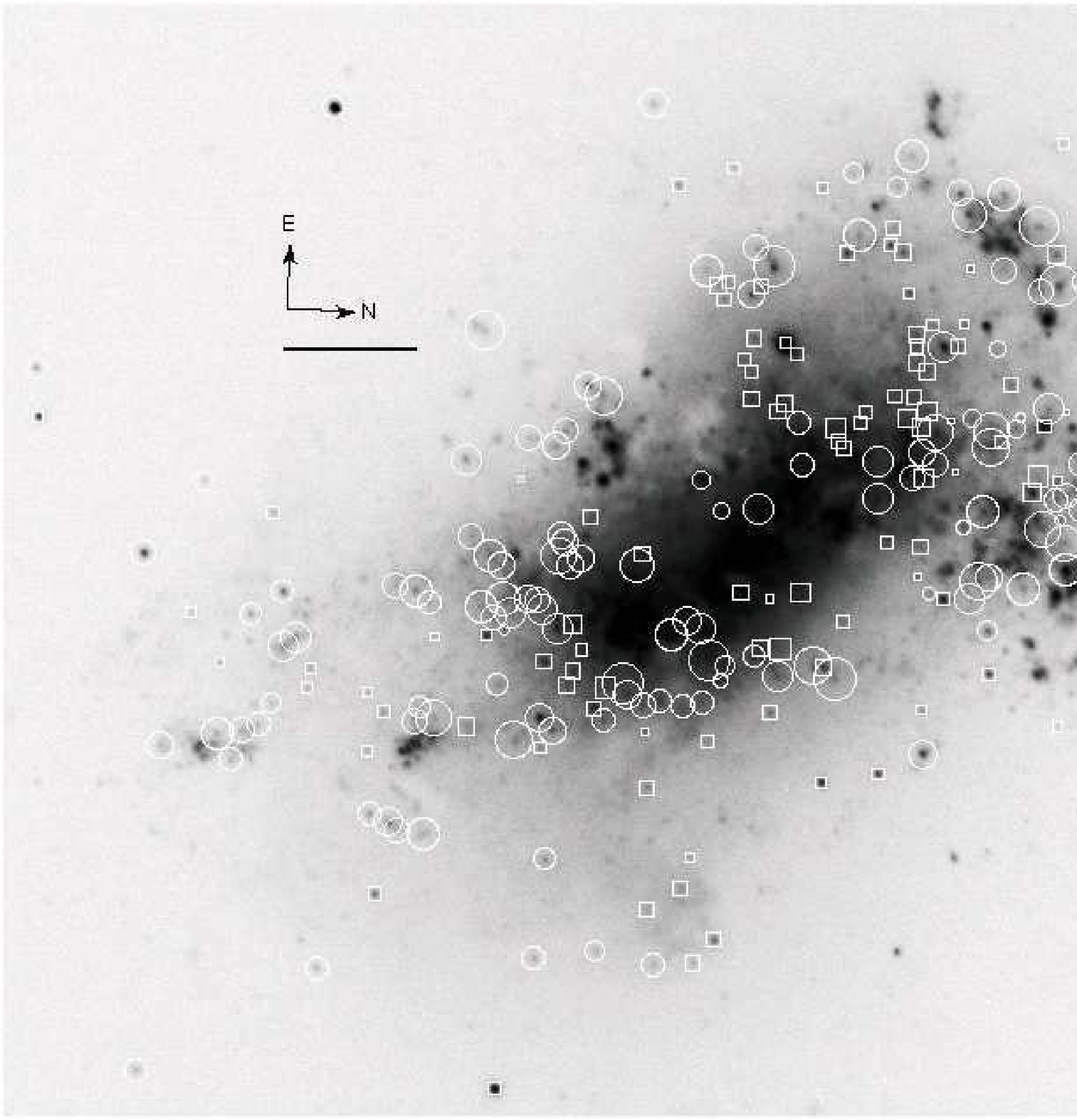}{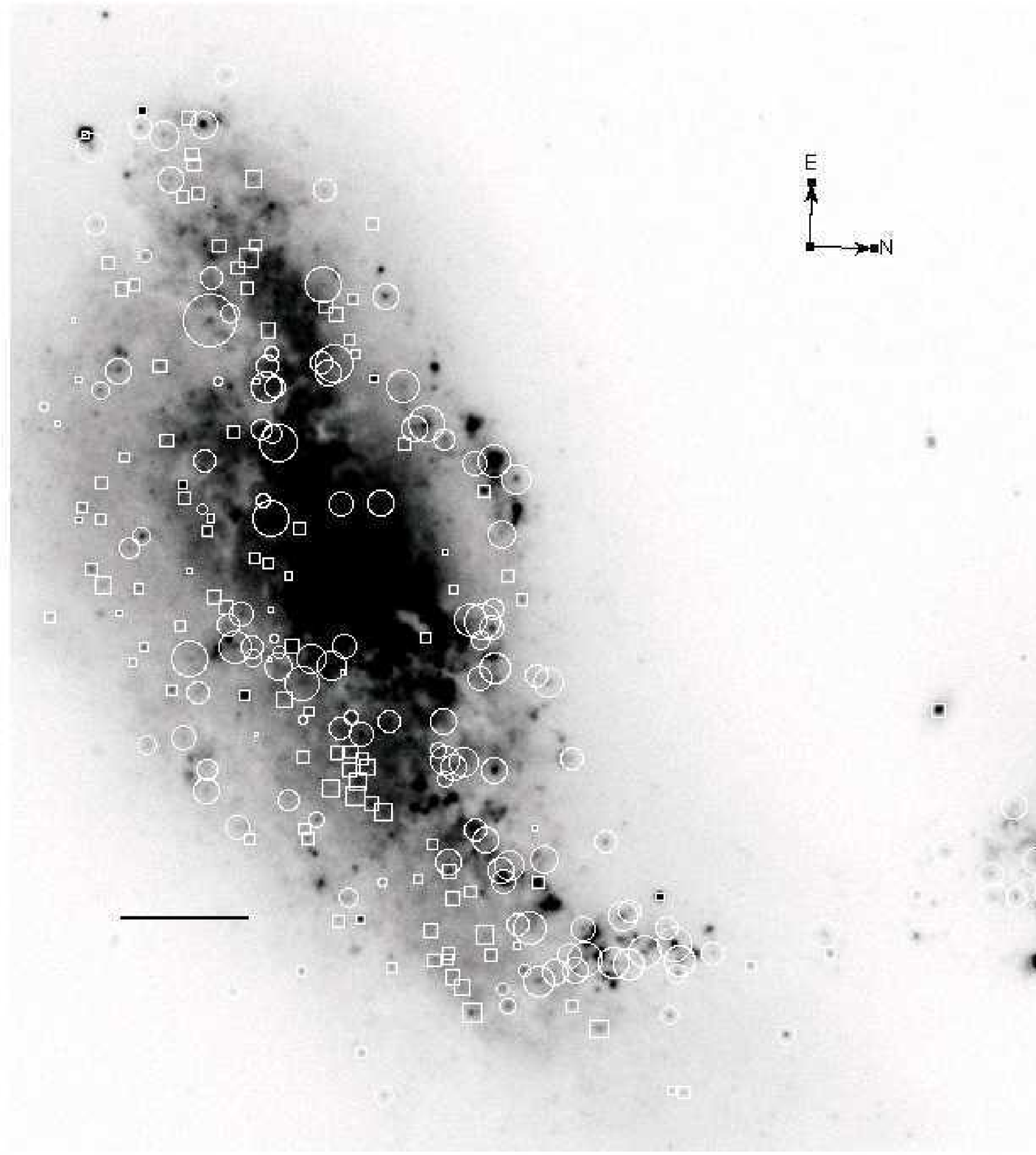}
	\caption{\label{fig1}$g'$ images of irregular galaxies NGC 2552 (upper left), NGC 3738 (upper right), NGC 4190 (center left), NGC 4214 (center right), NGC 4449 (lower left), and NGC 4485/4490 (lower right) with candidate clusters identified.  Circles indicate the location of young ($\leq 20$~Myr) clusters while squares indicate the location of older clusters.  In each case the size of the marker is proportional to the photometric aperture used.  The black bar shown in each image is 30$\arcsec$.}
	\end{figure}

\begin{figure}
	\plottwo{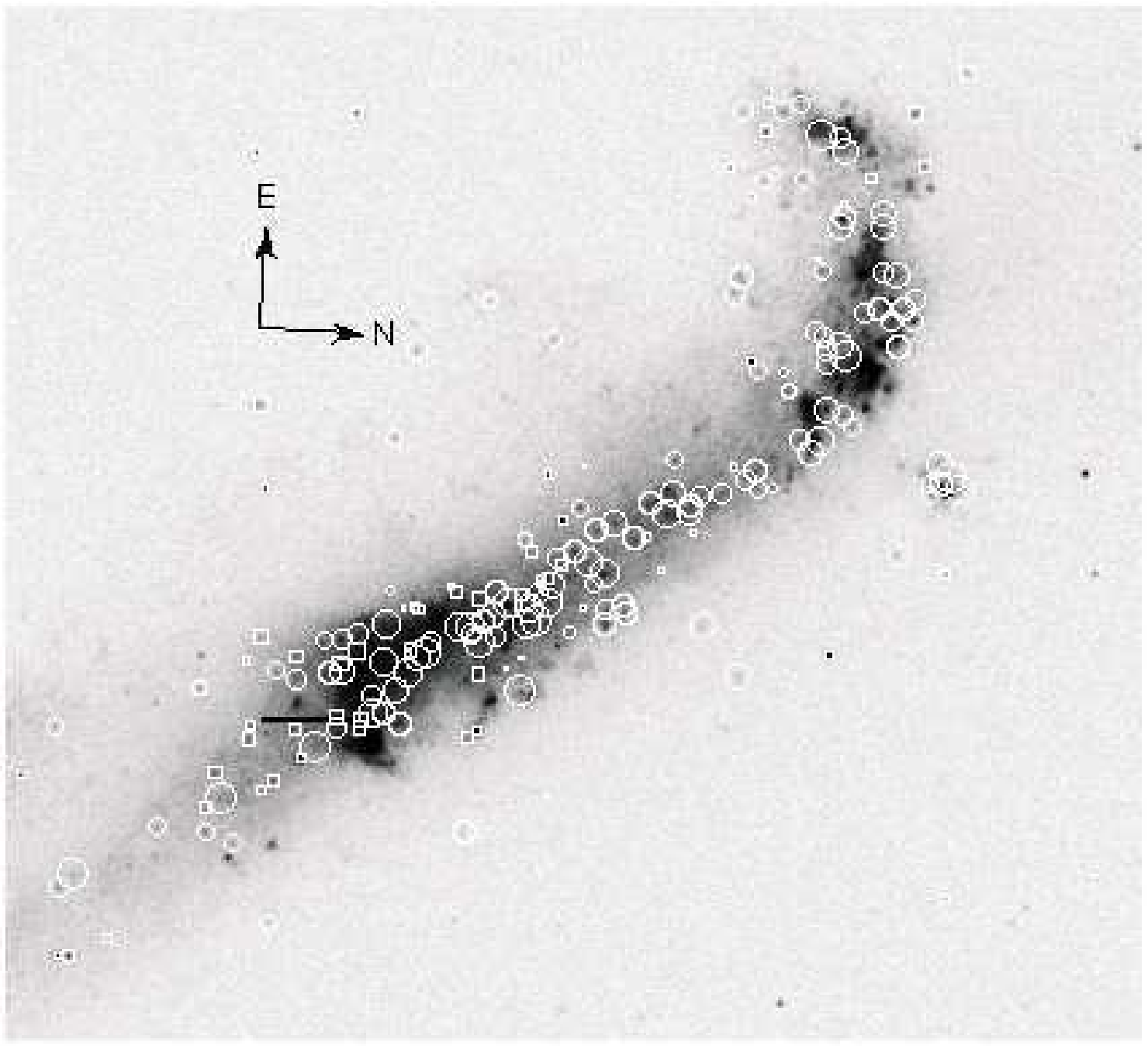}{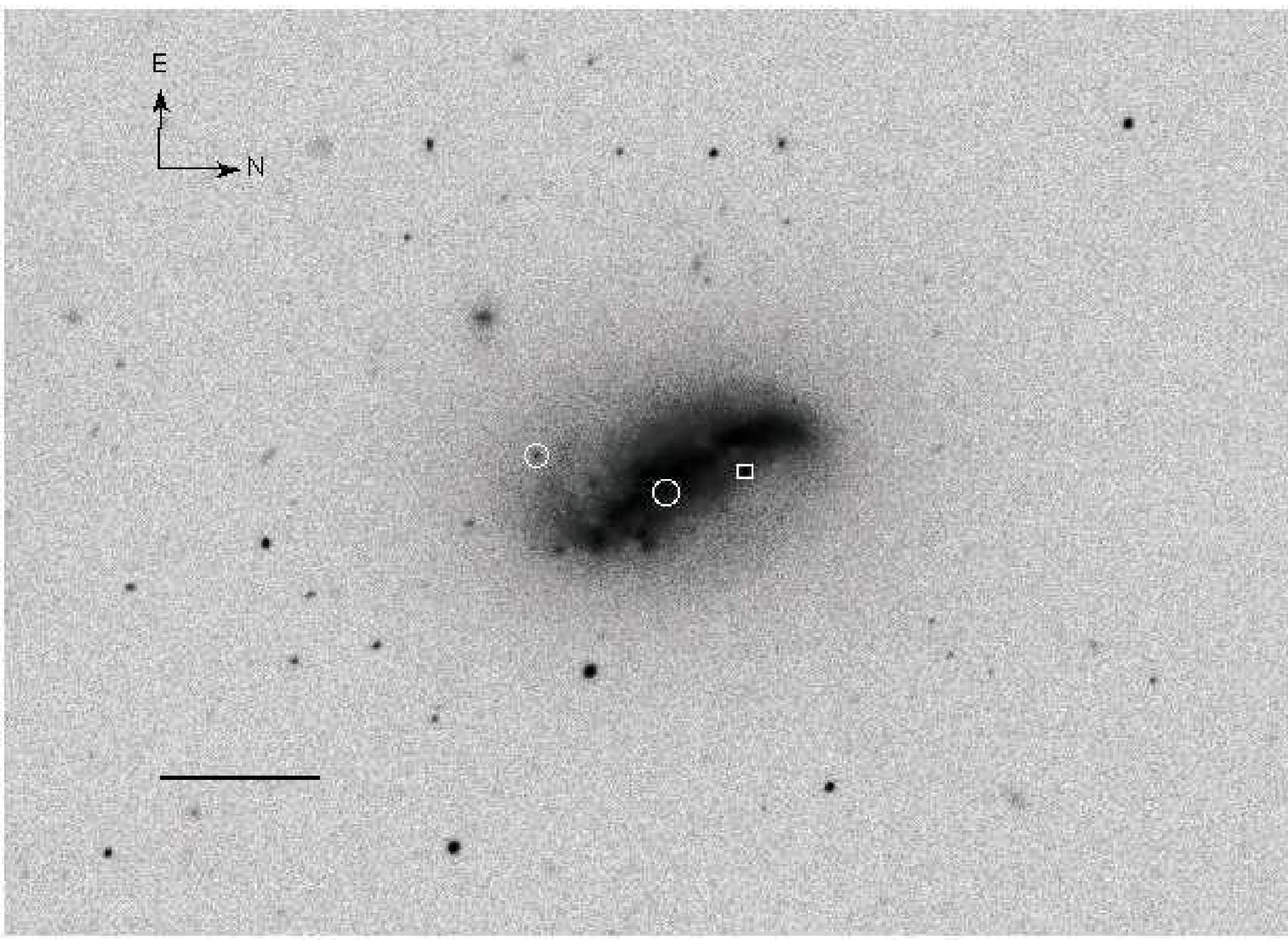}
	\plottwo{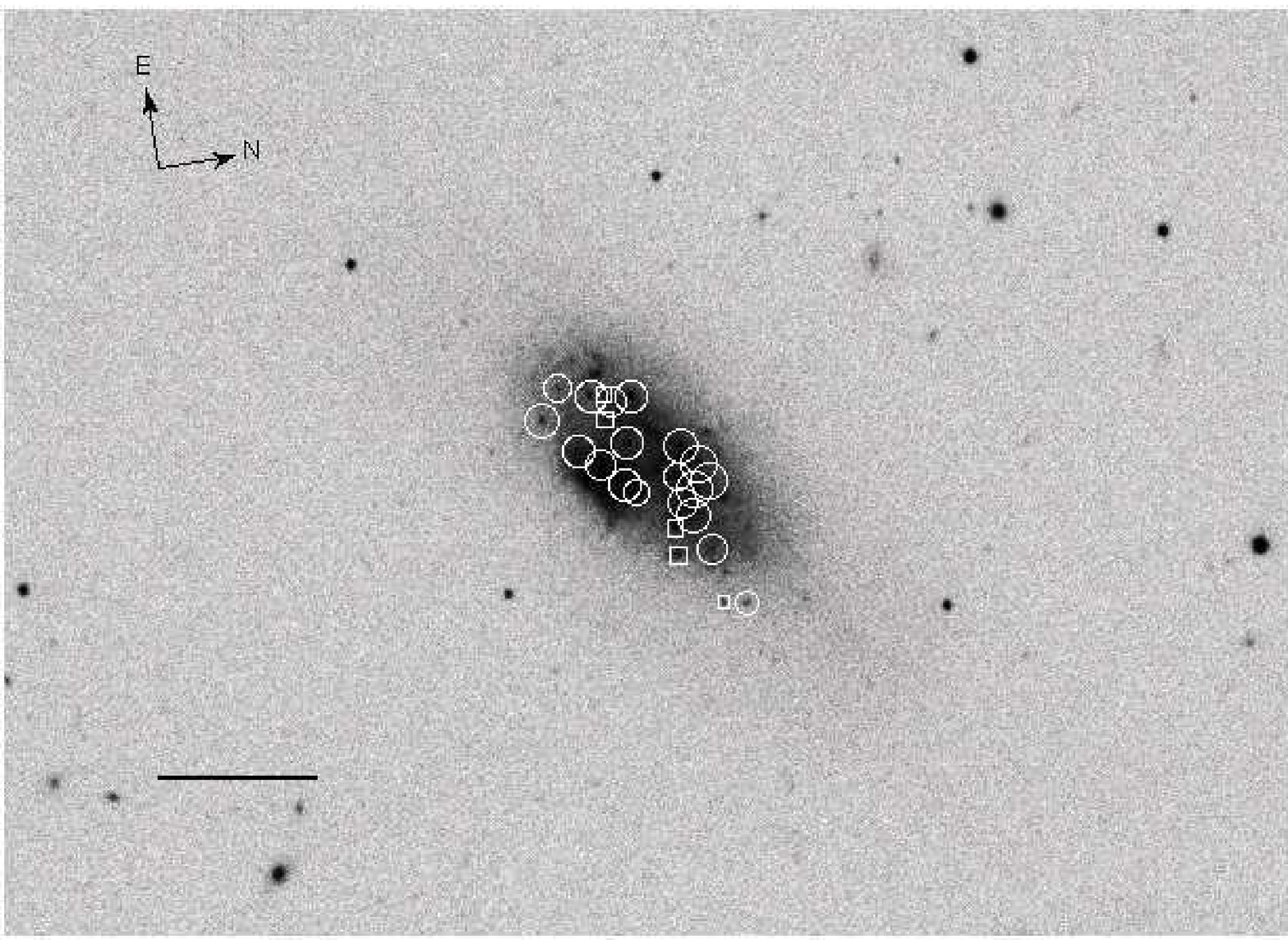}{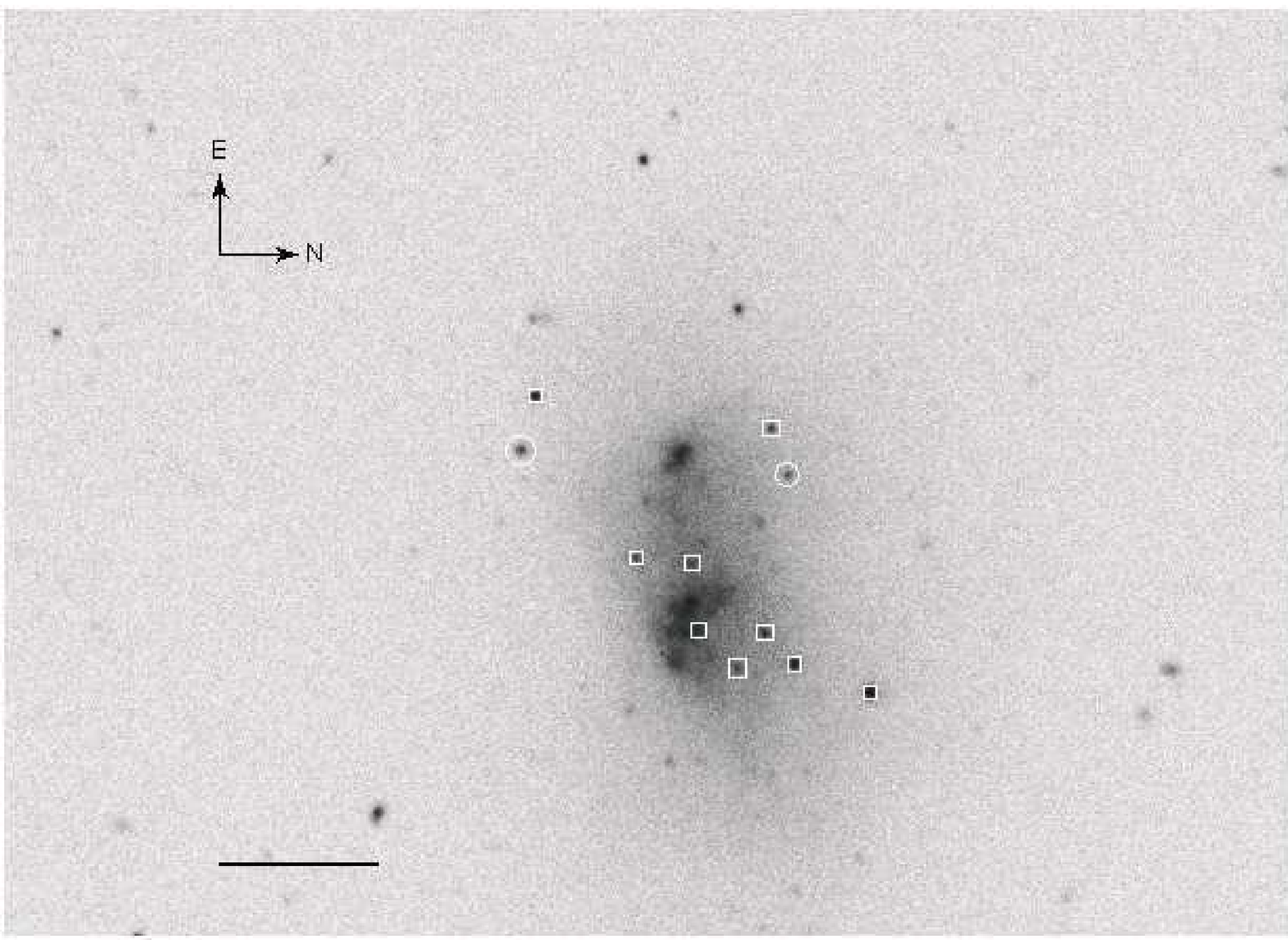}
	\plottwo{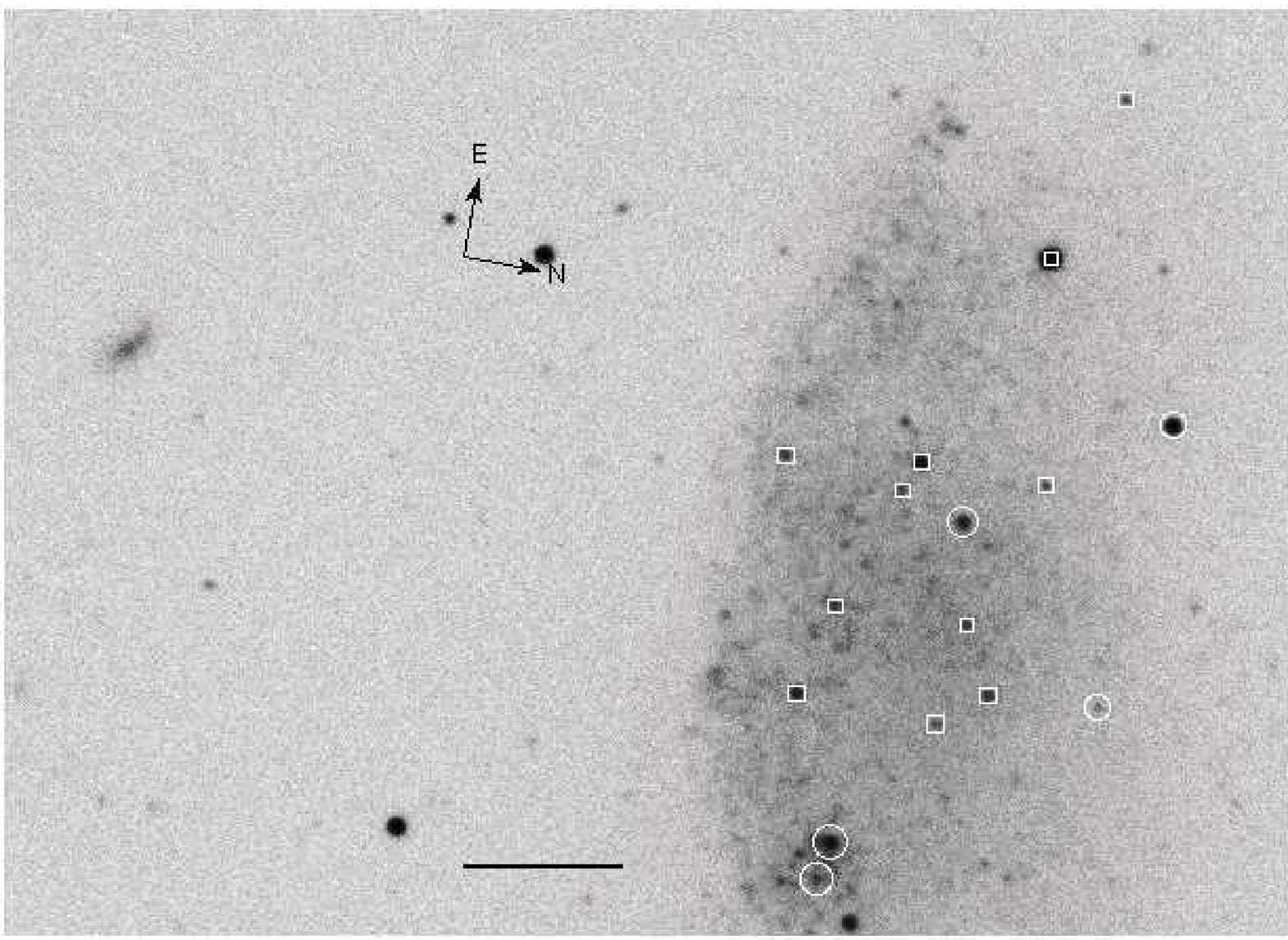}{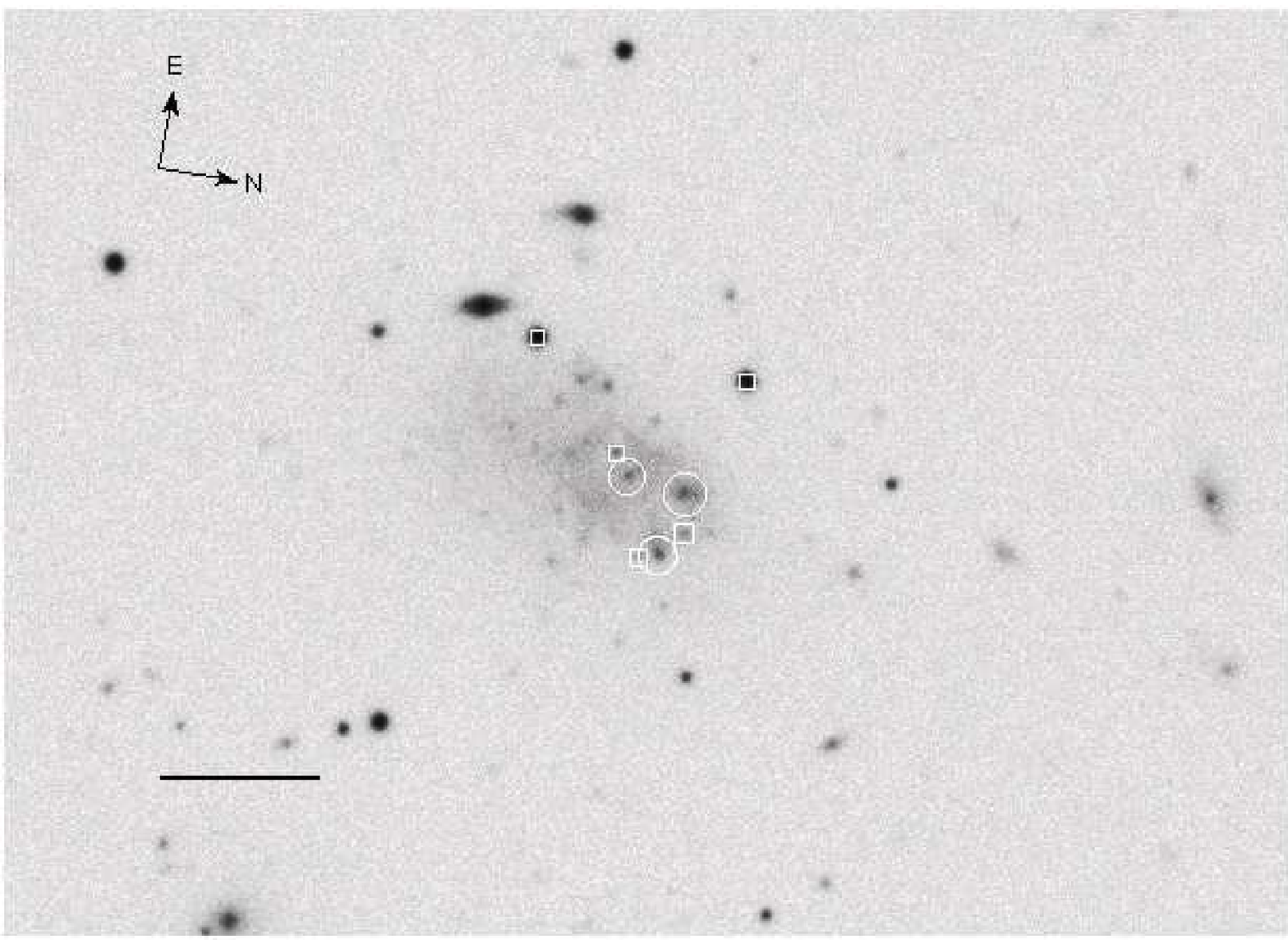}
	\caption{\label{fig1other}$g'$ images of NGC 4656 (upper left), IC 3521 (upper right), UGC 6541 (center left), UGC 7408 (center right), DDO 165 (lower left), and DDO 167 (lower right).  The labeling is the same as in Figure \ref{fig1}.}
	\end{figure}

\begin{figure}
	\plottwo{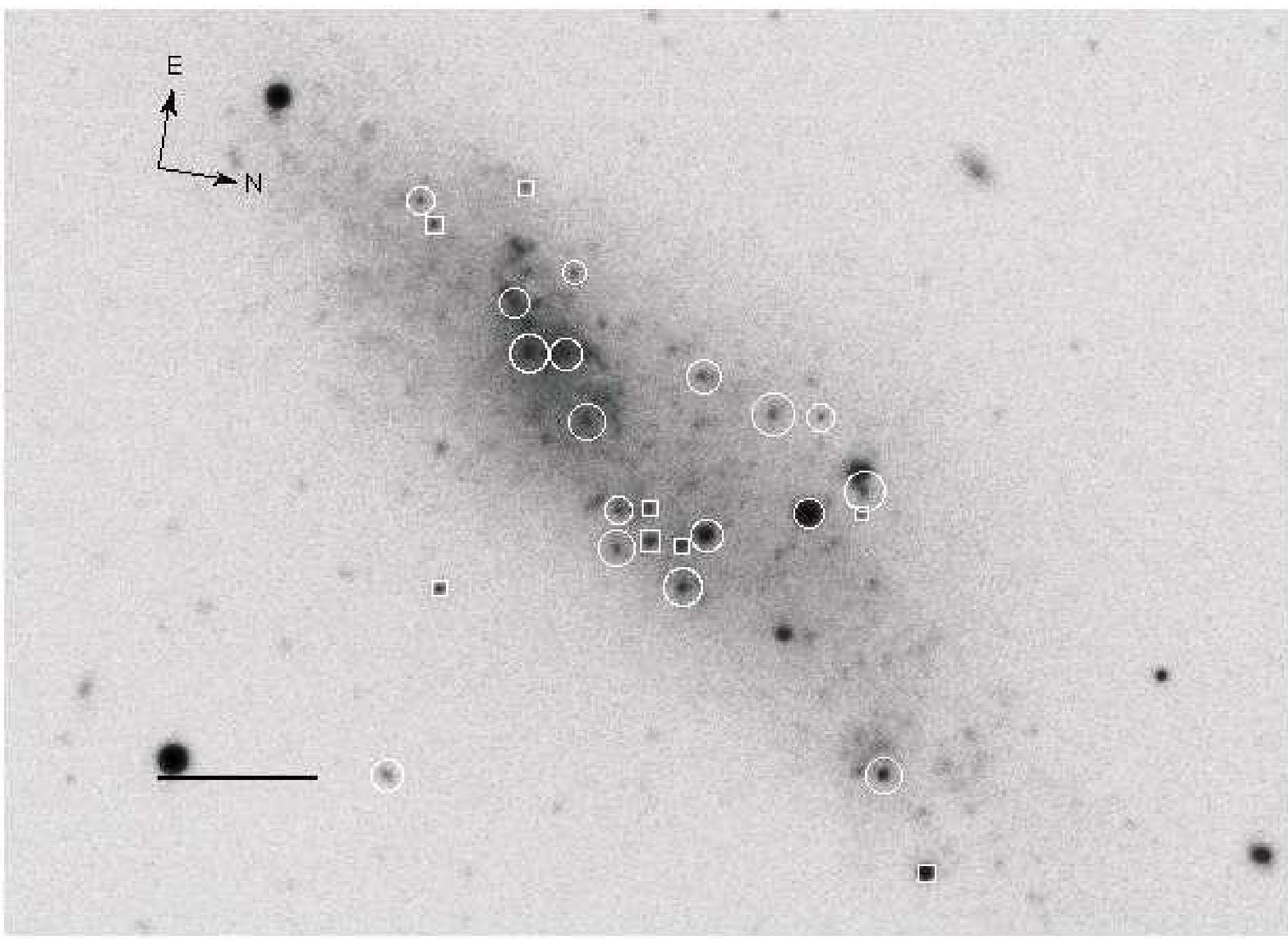}{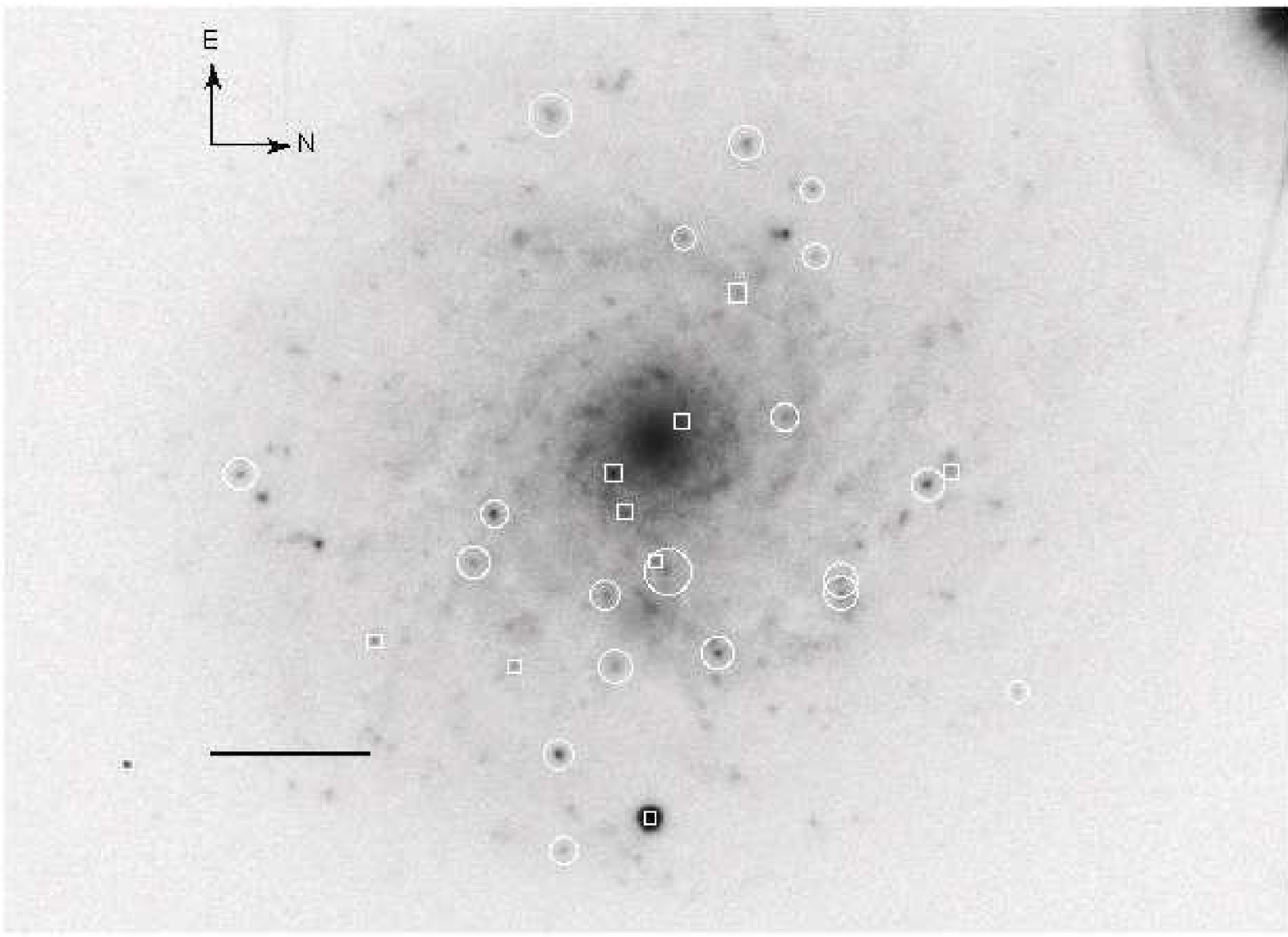}
	\plottwo{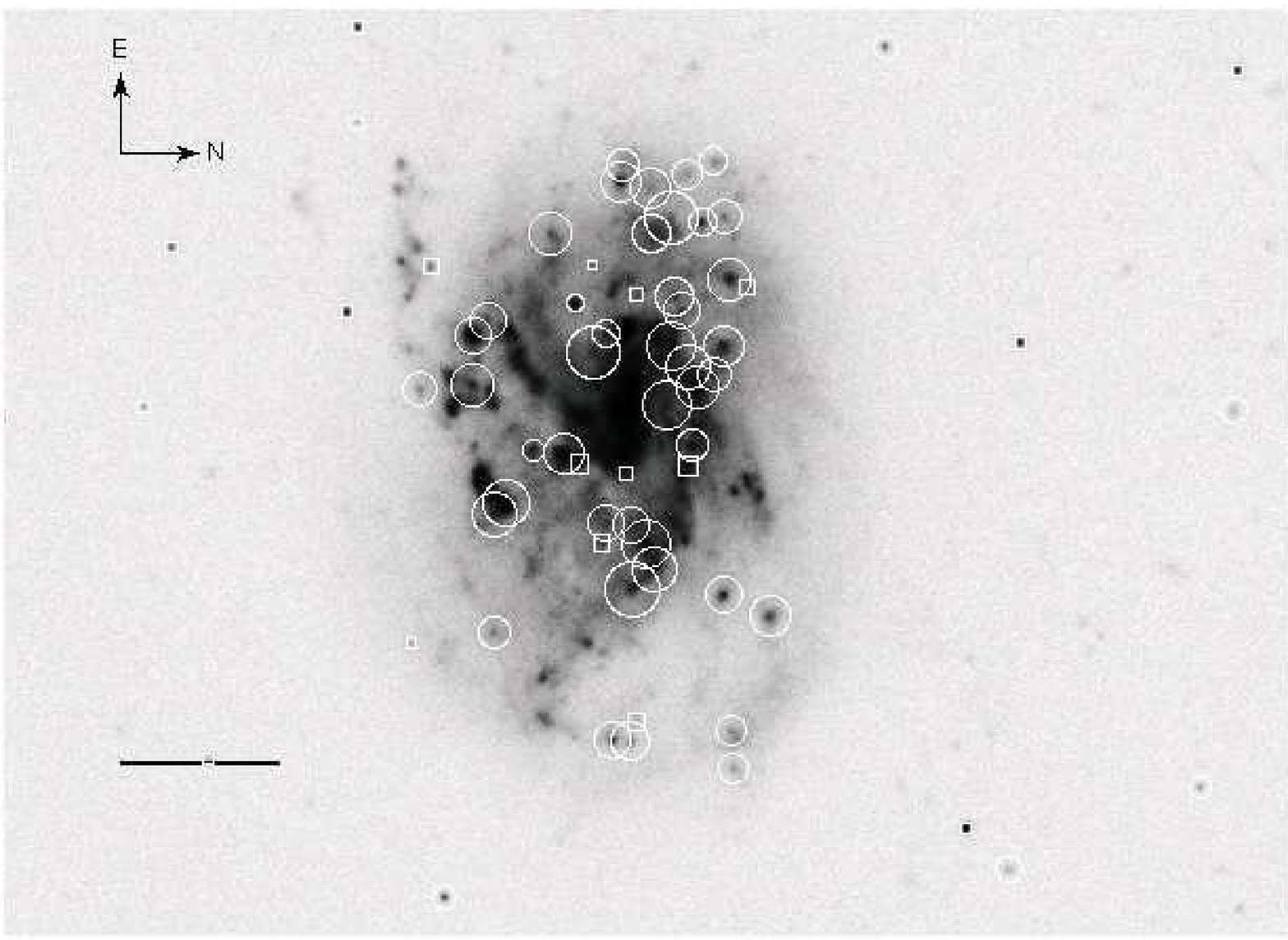}{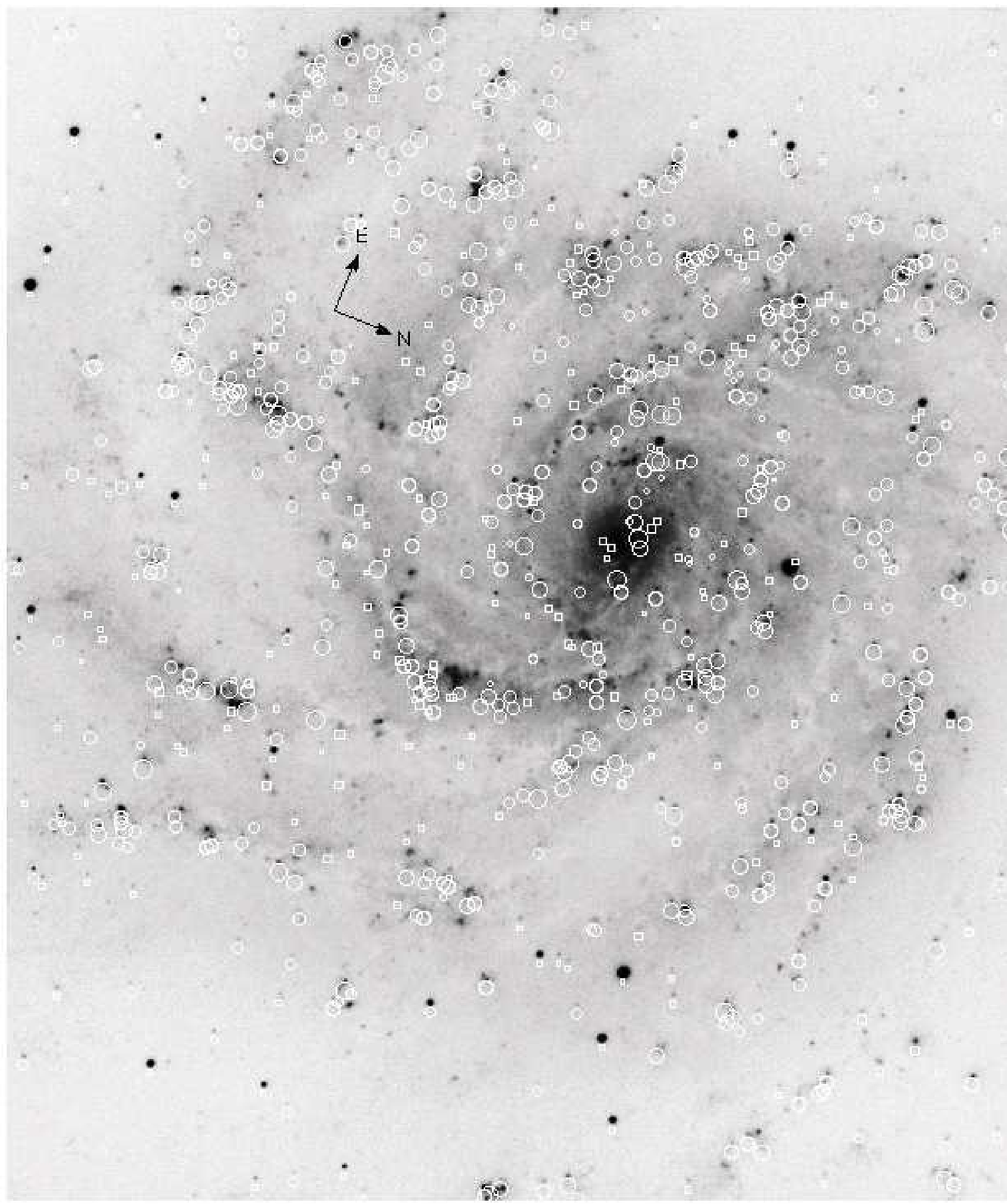}
	\caption{\label{fig1last}$g'$ image of DDO 168 (upper left).  Also shown are $g'$ images of spiral galaxies NGC 4571 (upper right), NGC 4713 (lower left), and NGC 5457 (lower right).  The labeling is the same as in Figure \ref{fig1}.}
	\end{figure}

The galaxies used in this study, listed in Table \ref{tab1}, came from
a search of the Third Reference Catalog of Bright Galaxies
\citep[hereafter RC3]{RC3}. The selection criteria for RC3 galaxies
were a $B$ magnitude $\leq$15, a Hubble stage number of $9.0-10.0$,
and a radial velocity $\leq$750~km~s$^{-1}$ (i.e., a distance
$\leq$10~Mpc assuming H$_0$=75~km~s$^{-1}$ Mpc$^{-1}$).  The list of
galaxies was then cross referenced to determine the inclusion in SDSS DR5 
\citep{SDSSDR5}, producing 13 galaxies, which defines our 
irregular galaxy sample.  SDSS DR5 $g'$ images of these galaxies are shown 
in Figures \ref{fig1}-\ref{fig1last}.

We also studied a small sample of three spiral galaxies with SDSS data
(NGC~4571, NGC~4713, NGC~5457), ranging between Hubble types Sb and
Sd, and distances from 5.8 to 9.5~Mpc (see Table \ref{tab1}).   SDSS DR5 $g'$ 
images of these 
galaxies are presented in Figure \ref{fig1last}.

\placetable{tab1}

\subsection{Photometry}

The only post-processing necessary for the SDSS images was to register
those images ($u'$, $i'$, and $z'$) with respect to the $g'$ image
coordinate system for each galaxy.
Clusters were located using the SExtractor automatic catalog
generation routine \citep{BA96} on the SDSS $g'$ image of each galaxy.
Young clusters are expected to be generally brightest in this band.
To aid in the location of potential clusters embedded in the diffuse
light the images were convolved with a Mexican-hat profile, with a FWHM
comparable to the stellar PSF. We selected objects with
signal-to-noise $\geq1.5$, chosen after some experimentation as the
optimum value to allow selection of relatively faint clusters but
without too many false detections.

Once the candidates were located, SExtractor was used to perform flux
measurements in all four bands using elliptical Kron apertures of
these objects.  During these measurements, as part of the automated
process, the influences of nearby objects were taken into account and
masked. The degree of crowding of the typical star cluster
environments in the irregular and spiral galaxy samples are
similar although there is substantial variation within each sample.

Once the fluxes and apertures for each potential cluster were returned, the list
of potential clusters was filtered to remove candidates with large photometric
errors because of apertures that extend too close to frame edges or
that overlap too much with adjacent sources. We also removed objects
with sizes $\geq1.5$ times that of the median stellar aperture, which
corresponds to a size of about 50~pc at a typical distance of 5~Mpc,
since these objects are likely to be blends of multiple clusters.

Next, the ratio of instrumental fluxes, $f$, to the reference flux, $f_0$, 
of cluster candidates were corrected for atmospheric extinction via
	\begin{equation}
	(f/f_{0})_{true} = \frac{f/f_{0}}{54 s}10^{-0.4(a + k \times m_a)},
	\end{equation}
where $a$ is the magnitude zero-point for the
given bandpass, $k$ is the extinction coefficient, and $m_a$ is the
airmass.  All parameters were obtained from the SDSS exposure parameter
file associated with each science image.
Once this correction has been made, the true
fluxes are converted to magnitudes on the inverse hyperbolic sine
magnitude system using
	\begin{equation}
	m = -\frac{2.5}{\ln(10)}\sinh^{-1}((f/f_{0})_{true}/(2b) + \ln(b)),
	\end{equation}
where $b$ is a band-dependent softening parameter \citep[see][for a
complete description]{LGS99}.  

For each irregular and spiral galaxy, local Galactic extinction was
removed using the value of $A_V$ obtained from the NASA/IPAC
Extragalactic Database (NED). We converted this extinction to the
\ugiz system using Table 6 of \citet{SFD98}.

Absolute $u'$, $i'$, and $z'$ magnitudes for each cluster were also
computed using distances from literature searches.  For all galaxies except NGC 4214, 
the distance was calculated from the radial velocities corrected for the
infall of the Local Group toward the Virgo cluster \citep{T02} and a
Hubble constant of 75 km s$^{-1}$ Mpc$^{-1}$.  For NGC 4214, the 
distance given in \citet{MCM02} was used.

\section{Determination of Extinction, Age, and Mass via the Color$^3$ Plot}
\subsection{Evolutionary Models}

Starburst99 version 5.0 \citep{S99} spectral energy distributions
(SEDs) were generated for an instantaneous burst model with $10^6
M_{\odot}$ total cluster mass, Salpeter IMF, upper and lower stellar
mass limits of $0.1\:M_\sun$ and $120\:M_\sun$, ages 0.01 Myr to 1 Gyr
in steps of 0.5 Myr, and metallicities of $Z=0.004$, $0.008$, and $0.02$.  
The model SEDs include contributions from both the stellar component and a 
nebular continuum \citep[see][for a description of the nebular conditions modeled]{S99}.  
The \ugiz magnitudes were generated from these model SEDs to derive the
extinctions, ages, and masses found below.  Model colors were used for
age determination while model absolute magnitudes were used for mass
determination.

\subsection{Overview of Color$^3$}

The color$^3$ (color versus color versus color) method that we introduce here is an improvement of
the age determination method of \citet{HEDM03}.  The method is
improved by using three broadband colors rather than two to decrease
the effects of color degeneracy encountered between 7 and 21 Myr, and
it also attempts to estimate the metallicity of the galaxy by
minimizing the global uncertainty in the derived reddening.  The
color$^3$ method consists of two routines: the dereddening routine and
the mass determination routine.  Internal reddening is determined by
shifting the cluster colors along an $R_V = 3.1$ reddening line in the
color-color-color plot until the clusters reach a minimum (least-squares) 
distance from the model curve.  The minimum distance was
determined by computing all possible reddening values between a
cluster's initial position and the bluest \ug and \gi model colors in increments
of 0.005 mag.  In the event that removing any additional reddening from 
the cluster results in it moving away from the model on the color-color-color 
diagram an $A_{V}$ of zero is adopted.  The length of the translation is then converted into an
\av for the star cluster, and reddening corrections are applied to the
absolute magnitudes.  

At this point, an age and its uncertainties can
be derived for the cluster by determining the model age corresponding
to the closest model point to the cluster.  The uncertainties of the
colors are first translated the same \av length along the reddening
line toward the model curve.  Next, the uncertainties are used to
define an ellipsoidal region.  This region is then divided into subregions
 where the central colors of each are compared to
the model in order to find their corresponding ages.  The greatest age
found is taken to be the upper limit while the least is taken to be
the lower limit.  After cluster ages and 
uncertainties were derived from the observed 
colors, the minimum age uncertainty of each cluster was set to the 
larger of 10\% of the age or 1~Myr, based on the expected systematic 
uncertainties in the star cluster models and the expected age spreads of 
massive star clusters \citep{TM04}.

This method is carried out on all three model metallicities, and the
results compared.  The comparison examines the number of clusters for
which ages have been determined and selects the model that results in
the largest number of clusters.  The metallicity associated with this
model is taken to be the metallicity of the galaxy.  This technique is
equivalent to minimizing the global error between the observed colors
and those predicted by the model.  In the event that two or more
models have the same number of clusters, the reddening dispersion is
then compared.  The model that corresponds to the minimum reddening
dispersion is taken to be the correct metallicity.  We expect the
derived reddening to be constrained across the galaxy
as many of the galaxies in our sample are face-on.  The galaxies being
face-on implies that reddening experienced by any given star cluster
should be dominated by the Galactic reddening in all but the youngest
clusters.  It should be noted that this method does not serve as a
rigorous determination of the metallicity of each galaxy.

\begin{figure}
	\plotone{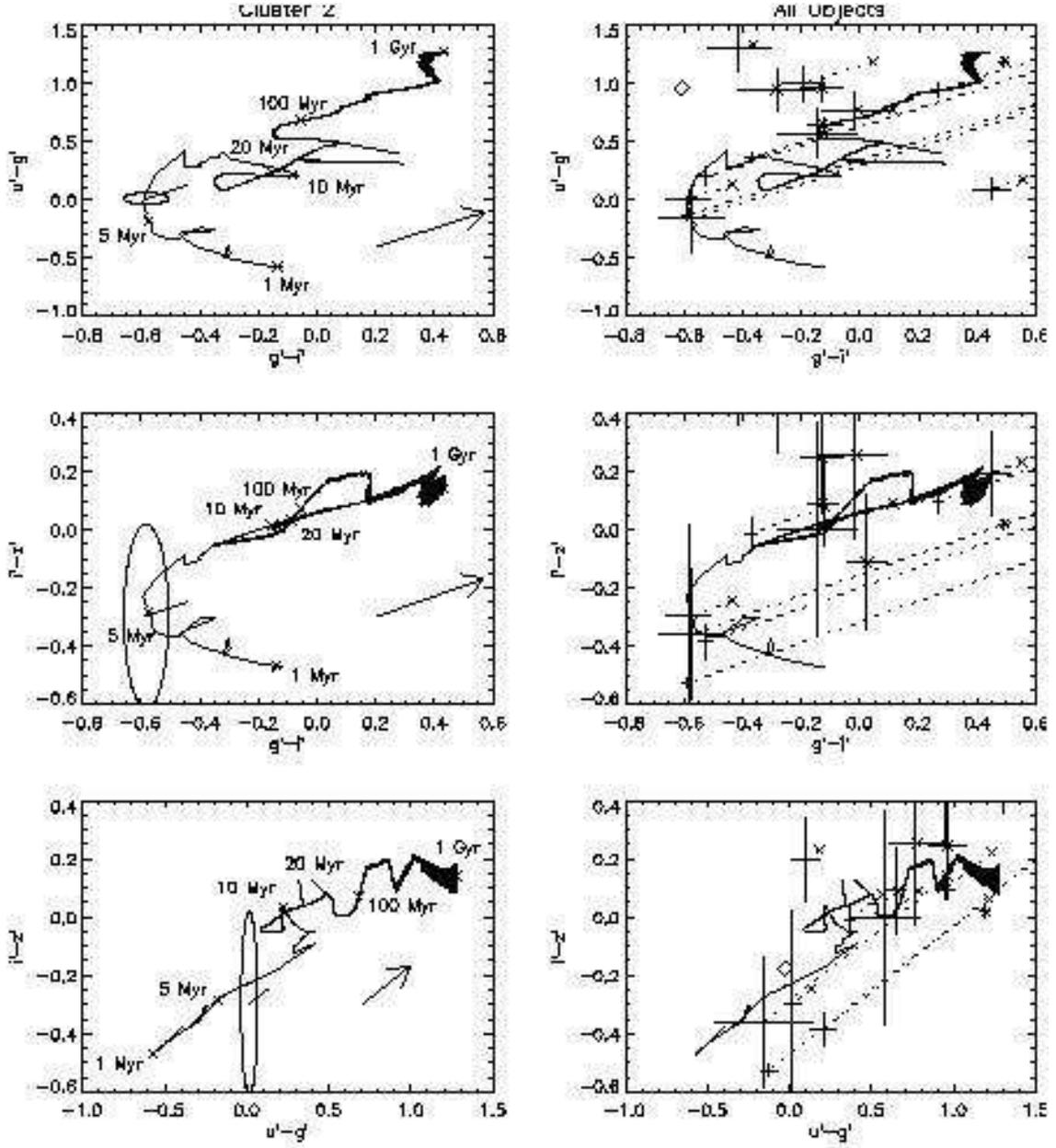}
	\caption{\label{fig2}Color$^3$ plots of \gi versus. \ug colors (top), \gi versus \iz colors (center), and \ug versus \iz colors (bottom).  Left Panel: this series demonstrates how color$^3$ derives an internal reddening of 0.28 mag and an age of $5.81^{+1.00}_{-1.00}$ Myr for cluster 2 in DDO 165.  The ellipses represent projections of the 1$\sigma$ uncertainty ellipsoid.  Ages along the solar metallicity model are indicated by crosses at 1, 5, 10, 20, 100, and 1,000 Myr.  The arrow in each plot represents reddening of A$_V$ = 1.00 mag.  Right Panel: the locations of dereddened clusters (plus signs) with 1$\sigma$ uncertainties for 15 clusters in DDO 165.  These clusters are connected to their initial (Galactic reddening corrected) locations (crosses) by dashed lines.  Also shown in this series is a cluster that was rejected by the color filters (diamonds).}
	\end{figure}

For this work, the three broadband colors used were \ug, \gi, and \iz.
The \ug and \gi colors were chosen to help differentiate between young
clusters and older stellar populations.  The \iz color was chosen
since these two filters are not significantly affected by emission
lines.  Figure \ref{fig2} shows an example of how this method derived the 
reddening and age of a $5.81^{+1.00}_{-1.00}$ Myr cluster in DDO 165 in 
addition to the dereddened colors of all clusters in DDO 165.

Once an age and its uncertainty has been determined, the mass
determination routine is called.  This routine uses an age and three
absolute magnitudes to determine a mass for the cluster via the ratio
of the observed flux in each of the three bandpasses to the model
fluxes.  The uncertainty in the mass is determined by using the
uncertainties in the absolute magnitudes as well as the uncertainty in
the age.  The absolute magnitudes $u'$, $i'$, and $z'$ were used in
this work for mass determination since $g'$ is more susceptible to
contamination due to nebular emission lines.

\subsection{Age Limit to Select Young Cluster}

We focus on young ($\leq 20$~Myr) clusters as these probe the initial cluster
 mass function most directly. From observations of clusters in
the Milky Way \citep{LL03} and other galaxies (e.g. \citealt{Fall05}), it 
appears as if a large fraction of clusters suffer from
``infant mortality'', i.e. they are only weakly gravitationally bound, if at all,
and suffer disruption so the mass function of older clusters is
significantly modified from the ICMF. The younger
clusters also have a higher light-to-mass ratio so a magnitude limited
survey can probe their mass function down to smaller masses than is
possible for older clusters.  The age limit of 20~Myr also counters the problem
 of heavy foreground Galactic stellar contamination in the $20$-$50$ Myr age range.  

Table \ref{tab2} contains the positions; absolute magnitudes; and derived 
reddenings, ages, and masses for the young clusters that comprise our sample.
Note that this table is available in full in the electronic version of the paper.  
The locations of these clusters in their host galaxies are circled in the $g'$ 
images found in Figures \ref{fig1}-\ref{fig1last}.

\subsection{Sources of Contamination}

Since we focused on the youngest star clusters, emission-line
contamination of the broadband colors was possible.  In addition,
since the star-formation histories of these galaxies are unknown, red
stellar populations from an earlier epoch may contaminate the star
cluster colors.  We removed those clusters from our ICMF analysis that
had apparent red stellar population contamination and emission line
contamination using the method described by \citet{HTI03}.  These
considerations were used to create the excluded color region needed by
the dereddening routine.  An additional filter was applied to remove
candidate clusters whose dereddened colors were greater than 3$\sigma$
and 0.50 mag away from the model colors.  This relatively high level
of 3$\sigma$ was chosen to counteract the color$^3$ method's tendency
to exclude clusters with small photometric uncertainties.  The 0.50
mag floor was added to address systematic uncertainties and includes
considerations for uncertainties in the metallicities, reddening, and
photometry.  Of the 2,184 potential star clusters identified in this
survey, only 48 (2.2\%) were rejected by the color filters or by being
3$\sigma$ and 0.5 mag away from the theoretical model colors.  Our
subsequent analysis will assume that there is no bias introduced by
this cut.

Another source of contamination is from foreground stars, background 
galaxies, supergiants, and stellar blends.  Using SDSS data helps limit 
the likelihood of imaging foreground stars by restricting the galaxy sample 
to high Galactic latitudes.  To determine the contamination level due to 
foreground stars and background galaxies we examined the fields around 
the 13 irregular galaxies.  Using SExtrator, we created a sample of field 
objects and applied the same selection criteria to them as we did the clusters.  
This field sample then had ages and masses derived using color$^3$.  This 
allowed us to estimate the sky density of these stars and galaxies to be, on 
average, 0.28 contaminants per square arcminute.  From this, we estimate 
that both the irregular and spiral galaxy samples are contaminated below 
the 10\% level.

Supergiants and stellar blends are dealt with using a mass cut-off of $10^{4.4}
M_\sun$. Hence, all clusters with masses below this level are expected
to have some degree of contamination.  This cut-off is roughly equivalent to 
excluding young clusters with $M_V > -9$.  Subsequent analysis in this
paper will ignore lower-mass clusters.

A summary of the effect on the sample sizes for the various rejection
methods used in this paper can be found in Table \ref{tab3}. It should
be noted that relatively large fractions of identified sources are
rejected at the various stages of this analysis. Most of these are
because of poor photometry (in any of the four bands), which can be due
to the clusters being relatively faint, in relatively crowded regions,
or extended (and thus likely to be blends of more than one
cluster).

\placetable{tab3}

\subsection{Sources of Systematic Error}

The variety of rejection methods introduced to deal with the
relatively poor angular resolution of the SDSS data (see \S2.2) and
the various sources of contamination (\S3.4) are potential sources of
systematic error.  From Table \ref{tab3} it is clear that the size of
the initial sample is affected most by rejection techniques that use a
combination of SExtractor quality flags and photometric aperture size.
The initial sample selection is strongly influenced by the angular
resolution, especially in regions where the diffuse galactic light is
prominent.  The detection method described in \S2.2 minimizes the
effect of diffuse light on source selection for the majority of the
star-forming area of each galaxy.  Detections in the nuclear regions,
however, are still limited due to a combination of strong galactic
light and the proximity of sources to each other.  This proximity
makes it difficult to estimate the flux contributions of the other
sources and increases the uncertainty in the fluxes to $>10$\%.  This
has a potential to influence the derived ICMF in two ways.  First, it
reduces the overall number of clusters that are used to determine the
ICMF and can increase the uncertainties in the power-law fits.
Second, some galaxies (e.g., NGC 4449, [\citealt{Boker01}]) have
numerous young clusters that are the result of a nuclear starburst.
In not being able to accurately measure clusters that may be part of
such a nuclear starburst, the initial sample may be skewed toward
older clusters where sufficient time has passed so that ``infant
mortality'' \citep{LL03}, over the first 20~Myr, has altered at
least the low-mass end of the ICMF.  In addition, the rate of nuclear
starburst activity may be different between the galaxies that compose
the irregular and spiral samples, thus leading to difficulties in
comparing the two derived ICMFs.

The photometric problem of many sources in close proximity to each
other also manifests itself in regions where potential clusters
themselves are highly clustered.  If young star clusters
preferentially form clustered around their parent giant molecular cloud (GMC), then excluding
these clusters would bias the selection method against the youngest
clusters.  Within each sample this could lead to the ICMF being
derived from a population of clusters that has already undergone some
evolution and disruption.  In addition, the degree of clustering in
irregular and spiral galaxies may be different due to the large
difference in galactic properties (Section 1).  This difference would make
direct comparisons between the two samples more difficult.

Subsequent analysis in this paper assumes that the potential
systematic errors in the sample selection method affect both the
irregular and spiral samples to the same extent.  Furthermore, we
assume that the potential systematic errors do not bias our derived
ICMFs in such a way that a direct comparison between the two samples
is inappropriate.

Systematic errors can also be introduced by the color$^3$ analysis
method both through limitations in the analysis approach and through
the low spatial resolution of the data.  \citet{Grijs05} reviewed a
variety of broadband imaging techniques, including a multi-parameter
minimization technique similar to the color$^3$ method used here, that
are used to derive star cluster parameters.  Using \textit{HST} data for
17 clusters in NGC 3310 and 20 clusters in the Antennae
galaxies they study the absolute systematic uncertainties associated
with these broadband methods.  In Tables A1-A10 of
\citet{Grijs05} they find that multi-parameter minimization techniques
systematically derive higher cluster masses than other methods.
However, they note that the relative age and mass distributions within
a population of clusters are robust across the various methods, i.e.,
observed features in either distribution are likely to be related to
features in the true distribution.  Thus, any systematic errors
introduced by the analysis method are not likely to influence the
overall shape of the derived ICMF.

\section{Derivation of the ICMF and Comparison Between Irregular and Spiral Galaxies}
\subsection{Completeness Corrections}

The completeness and sample bias corrections described by \citet{L99}
were applied to all galaxies with 20 or more detected clusters.  For 
these tests, 30 sets of randomly generated artificial clusters with masses 
between $\sim10^{4.4}$ and $10^{6.5}\:M_\sun$ were inserted into the $g'$ image 
of each galaxy.  The number  of artificial clusters was set to one-quarter 
the number of clusters found.  This dynamic  factor helps to reduce 
overcrowding in smaller galaxies while better sampling larger galaxies.  
The magnitudes used for these test clusters came from the appropriate 
metallicity Starburst99 models.  We limited the completeness and bias
corrections to clusters more massive than $\sim10^{4.4}\:M_\sun$ to
avoid contamination contributed by supergiants and stellar blends.
The artificial clusters used in this test were uniformly distributed
in age between 2 and 20 Myr.  Masses were sampled from a power-law
distribution with an index of -1.7.  The spatial distribution of
these clusters was determined by fitting an ellipse to the observed
cluster distribution and altering the semi-major axis until $\gtrsim$
90\% of the observed clusters were contained in this ellipse.  Once
these parameters of the ellipse were determined, the artificial
clusters were randomly distributed throughout its interior.  As a
final measure to insure that clusters generated were as
authentic as possible, random amounts of reddening were added to the
model magnitudes.  The reddening was uniformly distributed between an
$A_V$ of 0.02 mag and the median reddening magnitudes determined for
the particular galaxy.  At this point, the artificial clusters were
added to the images using IRAF's ``mkobjects'' routine.

Once the artificial clusters were added to the images, the SExtractor
routine was executed on the images using the parameters found in
\S2.1.  The returned object list was then compared with the list of
artificial clusters to determine the number of clusters recovered.
These results were then averaged across the various samples to arrive
at the completeness corrections presented in Table \ref{tab4} for both
the irregular and spiral samples.  For the lowest mass bin that we consider, 
$10^{4.40}-10^{4.60}\:M_\sun$, we estimate that our method of finding 
young clusters is 79\% complete for the irregular galaxy sample and 92\% 
complete for the spiral galaxy sample.  Higher angular resolution optical
searches by \textit{HST} would help us to probe regions of diffuse stellar light 
deeper and detect the less massive clusters with a higher detection rate.

\placetable{tab4}

\subsection{Variation of Cluster Frequency within the Galaxy Samples}
It should be noted that the frequency at which massive clusters appear
in our galaxy samples varies greatly.  From Table \ref{tab1} we see that three
galaxies dominate the irregular sample of massive
($>10^5\:M_\sun$) clusters: NGC 4449, NGC 4485/4490, and NGC 4656.  Using our
derived Hubble flow distances and the apparent \textit{B}-band magnitudes from
the RC3 yields absolute magnitudes of $-18.8$, $-17.5$, and $-18.7$,
respectively. This range of magnitudes places these galaxies at the
high end of the irregular galaxy luminosity distribution. Only one of
these galaxies, NGC 4656, has a classification other than ImB. Its
classification of SB/Sm(p) is likely due to its optical appearance
(see Figure \ref{fig1other}) being almost linear, with the possible remnant of a
spiral arm toward the northeast.

In the small spiral sample, NGC~5457 contributes about 90\% of the
massive clusters. Thus, our comparison of the mass functions of the two
galaxy samples is in essence a comparison of the clusters in NGC~5457
to those in NGC 4449, NGC 4485/4490, and NGC 4656. Larger samples of
galaxies are needed to improve upon this situation.

\subsection{The Initial Cluster Mass Functions}

\begin{figure}
	\epsscale{1.00}
	\plotone{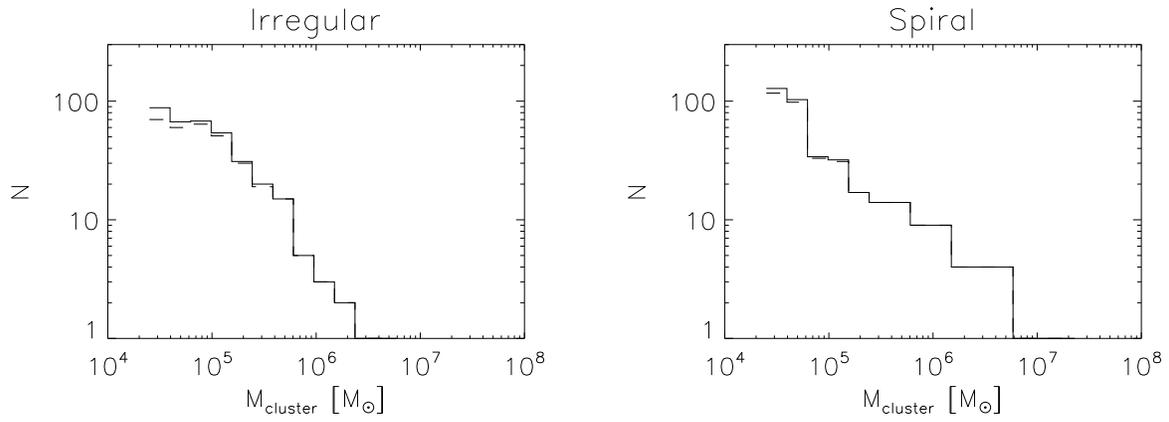}
	\caption{\label{fig3}Initial cluster mass function for irregular (left) and spiral galaxies (right), based on the mass histograms of young ($\leq 20$~Myr) clusters identified in these galaxy samples.  The uncorrected data are shown with dashed lines, while the completeness corrected data (\S 4.1) are shown with solid lines.}
	\end{figure}

The young ($\leq 20$~Myr) clusters were binned logarithmically by mass using the bin-sizing method described in \citet{Scott}.  This sizing procedure takes
into consideration the standard deviation of the data along with the
total number of data points to reduce any bias introduced by binning.
For our data set, this resulted in a bin size of 0.20 dex.  We applied
completeness corrections to each bin and assumed a combination of Poissonian 
uncertainty and uniform systematic uncertainties
of 30\% in the bin amplitudes.  Histograms for both samples are
presented in Figure \ref{fig3}.  In each case, the distribution of clusters spans
approximately the same range in mass.  Thus, despite being physically
smaller and possessing different internal conditions, irregulars are
capable of producing clusters with the same mass range as spirals.

\begin{figure}
	\plotone{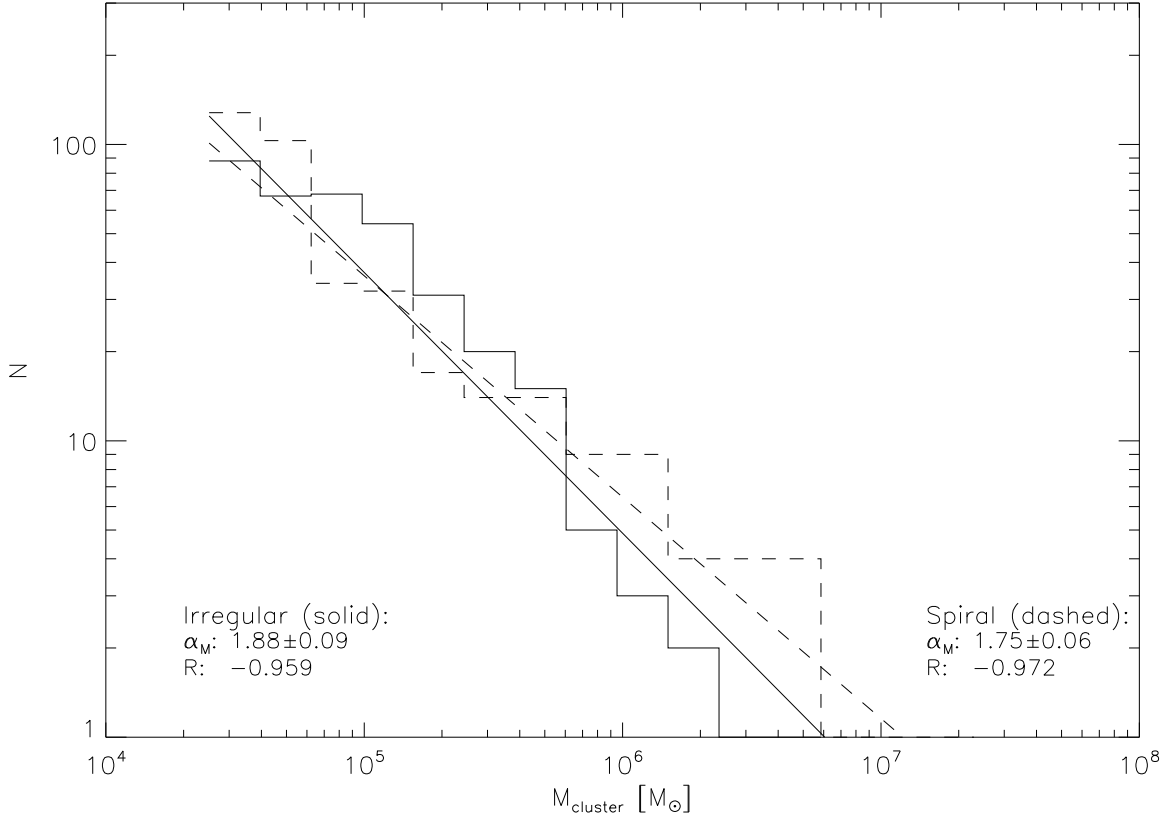}
	\caption{\label{fig4}Plots of the best power-law fits for the ICMFs in irregular (solid lines) and spiral galaxies (dashed lines).  Assuming the uncertainty in the bin amplitude is a combination of Poisson plus 30\% fixed systematic added in quadrature, the best fit power laws over the mass range $10^{4.4}$-$10^{7.5}\:M_\odot$ are shown with a solid line with power-law index $-1.88\pm 0.09$ for irregulars and a dashed line with a power-law index $-1.75\pm 0.06$ for spirals.  Statistically, these two distributions are indistinguishable.}
	\end{figure}

We fit power laws, $\frac{dN(M)}{dM} \varpropto M^{-\alpha_M}$, to the
ICMF in the range $10^5\:M_\odot<M<10^{7.5}\:M_\odot$, assuming errors
for each bin that are a combination of Poisson and a 30\% systematic
error added in quadrature. For the irregular ICMF, we derive
$\alpha_M = 1.88 \pm 0.09$ and for the spiral ICMF we derive $\alpha_M
= 1.75 \pm 0.06$.  The histograms along with linear regression fits are plotted in Figure \ref{fig4}.

The power-law indicies that we derive for these ICMFs are comparable to, though somewhat
shallower than, those derived by some other researchers. For example,
\citet{HEDM03} found an index of $-2.0$ to $-2.4$ for LMC and SMC clusters
in the mass range $\sim$10$^3-10^6\:M_{\odot}$. \citet{L02} found an
index of $\sim-2.0$ for clusters in disk galaxies. \citet{ZF99} found
the same index of $-2$ for young (${\la} 160$ Myr) clusters with masses
$10^4 \: \la $ M $\la \: 10^6\:M_\sun$ in the merging galaxies NGC
4038/4039. \citet{CVS04} found a index of $-1.56$ for young ($\la$ 20
Myr) clusters in the mass range $\sim10^4$-$10^6$ in the irregular
galaxy NGC 5253.

However, the absolute values of the power-law slopes are likely to be
influenced by the differing spatial resolutions of these studies. In
particular our analysis uses SDSS data with relatively poor spatial
resolution compared to, for example, \textit{HST} images so it is likely that
some of our more massive clusters are, in fact, blends of adjacent
smaller clusters, and this could affect the shape of the derived
ICMF. As an example of this kind of effect, in the Antennae galaxies
\citet{WS95} found a cluster LF with a power-law slope 
of $-1.8$ using pre-refurbished HST data while \citet{WZLFSM99} found 
a slope of $-2.1$ using higher resolution, post-refurbished \textit{HST} data.  
In a future study, we will examine this effect in a subset of our 
cluster samples using \textit{HST} data.

Another factor that could lead to systematic variations in the cluster
mass functions between different studies is the varying amounts of
spectral coverage of the observations, leading to different
uncertainties in the accuracy of derived cluster ages, reddening, and
masses. The SDSS data that we use here have the advantage of a relatively
broad spectral coverage and is homogeneous across our samples.

From our results, there is some evidence from the lowest-mass bins
($M<10^5\:M_\odot$) in both the spiral and irregular
samples that the single power-law description may break down at these
masses, even after completeness corrections have been made. We detect
relatively few clusters in the mass range
$10^{4.4} M_\odot\:<$ M $<\:10^5M_\odot$, and this contrasts with
some previous studies (e.g., \citealt{HEDM03} study of the LMC and
SMC). It is possible that our completeness correction has been
underestimated; for example, some of the filtering that we carry out for
the cluster sample based on colors and nebular line contamination may
tend to remove lower-mass clusters, and this possible bias is hard to
quantify. Another possibility is that, again, the relatively poor spatial
resolution of our galaxy images may cause us to group together
associations of several smaller clusters into a single object, which
we treat as a ``cluster''. The size of the point-spread function is
several tens of parsecs at the typical distance of galaxies in our
sample. This hypothesis could be tested by seeing if the young
association (i.e. groups of young stars within $\sim50$~pc of each
other) mass functions in the LMC, SMC, and nearby spirals show a break
at $\sim10^5\:M_\odot$.

Our ICMF power-law index of $-1.88\pm0.09$ for irregular
galaxies over the mass range $10^{4.4}$-$10^{7.5}\:M_{\odot}$ is
consistent within the uncertainties to that derived from the spiral
galaxy sample, i.e. $-1.75\pm0.06$.

We have also compared the spiral and irregular raw ICMFs
(i.e. with no completeness corrections applied) for masses above
$10^{4.4}\:M_\sun$ using a Kolmogorov-Smirnov test. There are 321 such
clusters in the irregular galaxies and 358 in the spirals.  This
test yields a \textit{p}-value of 0.633, indicating that the distributions are
statistically indistinguishable.

\section{Discussion: Implications for Star-Formation 
Theories}\label{S:discussion}

\subsection{Effect of Galactic Shear}

The ICMFs in dwarf and spiral galaxies appear
to be very similar in spite of the relatively major differences
between these galaxy types.

Irregular galaxies tend to have rotation
curves that are rising with increasing galactocentric radius. This
contrasts with the more nearly flat rotation curves of typical spiral
galaxies, such as those making up our sample of spirals. In particular
NGC~5457 is classed with a Hubble stage of 6.0 and has an extended
Hubble type of SABcd(rs), indicating it is a mixed bar and non-barred
type, somewhere in between Sc and Sd in terms of the winding of the
arms, and the internal organization in the galaxy consists of a
mixture between an inner ring and spiral arms. The rotation curve of
this galaxy is not as flat as a classic, earlier-type spiral, i.e. it
shows a sharp increase within the innermost 1 kpc, then is flat out to
$\sim$5 kpc, then increases again, peaking at about 250~$\rm
km\:s^{-1}$ around 8 kpc \citep{M101}. However, it should be noted that we
do not generally have detailed information on the rotation curves of
the galaxies in our irregular and spiral samples. A flat rotation
curve means that gas experiences large shearing motions due to
differential rotation in the disk. This enhances the collision rate
between GMCs \citep{tan00}. Shear is absent in a rotation curve that
rises linearly with radius (i.e. solid body rotation). The observed
similarity of the ICMFs in irregular and spiral galaxies implies that
the process determining the masses of clusters does not depend on
galactic shear and is, thus, probably operating on scales smaller than
the tidal radius of giant molecular clouds in a typical flat rotation
curve spiral, i.e. $\la 100$~pc. \citet{ZFW01} and \citet{WGLFCBSZM05} 
found a similar lack of influence of shearing velocity gradients on 
cluster formation in the Antennae galaxies.  

According to the model for star-cluster formation by \citet{EE97}, the
mass function of clusters reflects that of unstable interstellar gas
clouds, and a universal cluster mass function from galaxy to galaxy
reflects the universal nature of turbulence, at least on scales
relevant to star cluster formation. Our results are consistent with
this universal model.

One explanation for the global star formation rates in galaxies
\citep{K98} is the triggering of star formation by cloud
collisions \citep{tan00}, the rate of which depends on the galactic shear
rate. The main prediction of this model is a smaller star-formation
efficiency per orbital time in irregular galaxies compared to
spirals. The data presented in this paper can be used to estimate
global star-formation rates in these galaxies. With data on the gas
content of the star-forming regions of these galaxies (e.g. \citealt{LBSB05}), 
this question will be able to be addressed.

\subsection{Effect of Metallicity}

Irregular galaxies, being typically of smaller
mass and with a higher gas fraction than normal spiral galaxies,
tend to have much lower metallicities. This does not appear to have
any bearing on the ICMF.  Note that the range in metallicity of the
Starburst99 models considered, $Z\sim0.004$-$0.02$, is not an accurate
assessment of the range in metallicities of our galaxy samples, and
the metallicity recorded in Table~2 for each galaxy is a somewhat
arbitrary choice intended to maximize the number of clusters found by
the color$^3$ method (see \S3.2). Good measurements of the metallicities
of the galaxies in our samples are not generally available, so a more
quantitative discussion of the range of metallicity over which the
ICMF is invariant is not possible at this stage.

One mechanism by which metallicity might be expected to affect star
formation involves the degree of ionization and heating inside
molecular clouds due to far-UV photons. In a higher metallicity cloud
with a higher dust-to-gas ratio and higher mean extinction, the far-UV
photons do not penetrate as easily. If magnetic fields are important
for preventing GMC collapse and thus regulating star formation
\citep{MCK89} and if the creation of magnetically super-critical
regions occurs primarily via ambipolar diffusion in regions of low
ionization, then one expects such regions to require smaller amounts
of shielding gas in a higher metallicity galaxy. In other
words, shielding a given mass of gas in a low-metallicity galaxy
requires it to be in the center of a more massive GMC, i.e. with a
larger total mass surface density, $\Sigma$, than in a
high-metallicity galaxy. The pressure due to the self-gravity of the
gas will be of order $\simeq G\Sigma^2$, and the combination of higher
pressure and more surrounding material may lead to higher star-formation efficiencies and higher-mass clusters. See \citet{TM04} 
for a more extensive discussion of feedback processes in
forming SSCs. Since we do not observe such a difference
in the ICMFs of (presumably lower metallicity) irregulars and
(presumably higher metallicity) spirals, we conclude that either
magnetically super-critical regions are created by turbulent diffusion
of field strength rather than ambipolar diffusion or that other
processes, such as turbulent fragmentation, set the ICMF on scales
smaller than the super-critical regions of GMCs.

\subsection{Continuity of the ICMF}

Since the derived star cluster masses presented for the irregular
galaxy sample cover a wide range of masses, we are able to examine the
question of the continuity of the ICMF discussed in \citet{BHE02}.
The significant linear correlation \textit{p}-value ($<0.01$) for the power-law fits implies that the ICMF is well modeled by a single power law
in the $\sim 10^{4.4}$-$10^{7.5}\:M_\sun$ mass range. We conclude that the
ICMF in our sample of irregulars is well described by a
continuous power law.
Due to completeness and
stellar contamination issues in the lower mass bins, the only
conclusion that can be drawn about the issue of continuity is that if
a break in the power law exists, it occurs for clusters with masses
$<3 \times 10^{4} M_\sun$.

It should be noted that a power-law index of $\alpha_M<2$ implies that
the total star formation occurring in clusters (in a given mass range)
is dominated by that in the largest clusters. If the ICMF does not
significantly steepen below $5\times 10^4\:M_\odot$, then our results
imply that the star formation in irregular and spiral galaxies
is dominated by that occurring in SSCs or super
associations.  However, the derived power-law indicies of our ICMFs are 
close to $-2$, and at this value each logarithmic interval in cluster mass 
contributes equally to the total mass of new stars.

\subsection{Evolution of Extinction}

\begin{figure}
	\plotone{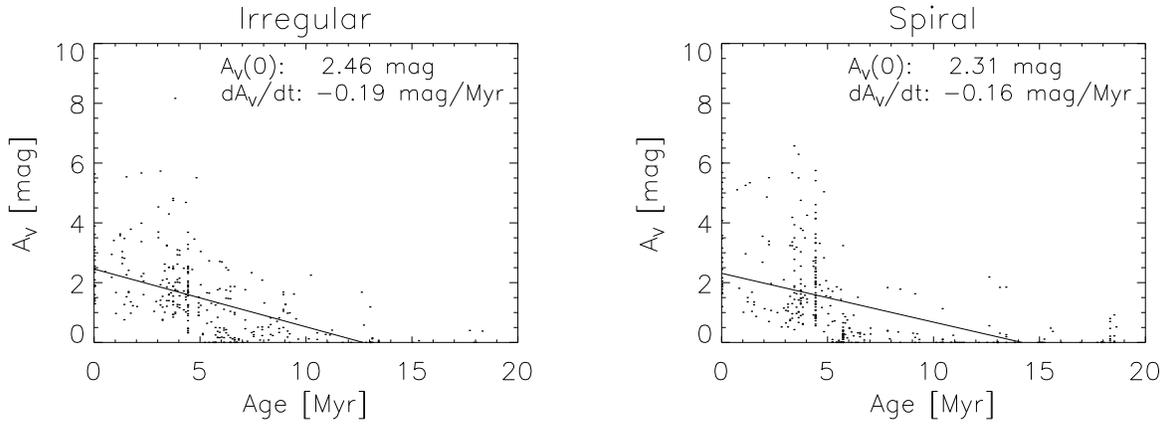}
	\caption{\label{fig5}Visual extinction versus age of young massive clusters in irregular (left) and spiral (right) galaxies. We see an apparent evolution in the extinction of clusters as they age. Clusters younger than 5 Myr always have at least a few tenths of a magnitude of local \textit{V}-band extinction. Clusters with ages 5 Myr or older have a $\gtrsim$50\% chance of being relatively free ($A_V<0.1$~mag) of extinction. This 5-10~Myr timescale is thus the characteristic time for massive clusters to destroy or move away from their natal dust clouds. Note that some of the relatively high values of extinction at later ages are likely caused by our line-of-sight intercepting local, but unrelated, clouds in the host galaxies.  The concentration of clusters at particular ages is due to extremities of the tracks of model cluster evolution in color space, to which the color$^3$ method concentrates clusters with relatively large photometric uncertainties by adjusting the amount of extinction. Thus some of the scatter in the derived extinction is also likely to be due to a combination of photometric errors coupled with systematic uncertainties in the positions of the tracks of model clusters in color space. Nevertheless, the evolution of the mean cluster extinction does appear to show a real decrease to very small values over about 5-10 Myr.}
	\end{figure}

We expect that younger clusters should suffer from a greater
amount of local extinction relative to older clusters. To test this
hypothesis, we examined how the local extinction estimate for our
clusters varied with cluster age. In Figure \ref{fig5}, we see an
apparent decrease in the cluster extinction with age.  We also note
that there are no clusters younger than 5~Myr that have an $A_V$ less than
0.1~mag while 15\% of clusters in irregulars and 28\% of clusters in
spirals older than 5~Myr have an $A_V$ less than 0.1~mag. 
Performing a linear fit between the extinctions and ages we derive
that an average cluster in an irregular galaxy has $A_V$ decrease by
0.19 mag Myr$^{-1}$, while an average cluster in a spiral galaxy has
$A_V$ decrease by 0.16 mag Myr$^{-1}$.

The above results imply that 5-10~Myr is the characteristic timescale for a
cluster to destroy or move away from its natal molecular cloud, which
is consistent with estimates of this timescale based on studies of
young Galactic star clusters \citep{LBT89}.  \citet{WZ02} and \citet{MLTG05} 
have found similar results from extragalactic studies of massive young clusters.

We examined the mass dependence of these relations, defining a
``low-mass'' sample from $10^{4.4}$ to $10^5\:M_\odot$ and a
``high-mass'' sample from $10^6$ to $10^{7.5}\:M_\odot$. The high-mass
clusters in irregular and spiral galaxies start with high values
of mean $A_V$ at zero age of 5.29 and 5.41~mag, respectively. The
corresponding values for the low-mass clusters are 1.95 and
1.44~mag. The rates of decrease of $A_V$ are -0.29 and 
-0.33~mag~Myr$^{-1}$ for the high-mass clusters in irregular and spiral galaxies,
respectively, while the rates are $-0.14$ and $-0.10$~mag~Myr$^{-1}$ for the
low-mass clusters, respectively. Thus the 5-10~Myr timescale for
reduction in extinction is similar for all these sub-samples.

\section{Summary}

We have presented the age and masses determined for 321
young ($\leq 20$~Myr) star clusters in 13 irregular
galaxies and 358 young star clusters in three spiral galaxies based
on an automated analysis of SDSS data.  From this, we
have found the irregular ICMF to be well fit by a power law
$\frac{dN(M)}{dM} \varpropto M^{-\alpha_M}$ with $\alpha_M = 1.88\pm0.09$ for 
clusters over the $10^{4.4}$-$10^{7.5}\:M_{\odot}$
mass range. The equivalent index for the spiral ICMF was found to be
$\alpha_M = 1.75\pm0.06$. The irregular and spiral ICMFs are
statistically indistinguishable in spite of expected differences in 
galactic shear and metallicity.

Our derived ICMFs are slightly top-heavy compared to some previous
studies, perhaps because of the generally poorer spatial resolution of
the SDSS data; it is likely that many of our more massive ``clusters''
are blends of adjacent smaller clusters. Nevertheless, this does not
affect our conclusion on the similarity of the irregular and
spiral ICMFs since these data were analyzed in the same way and share
the same systematic errors.

We also find a characteristic timescale of about 5-10~Myr for clusters to move
away from or destroy their natal gas and dust clouds.

\acknowledgments We thank the anonymous referee for detailed comments that 
led to a significantly improved paper. We also thank Bruce Elmegreen, Mike Fall, 
Chris McKee, and Brad Whitmore for comments.  This research has made use of the NASA/IPAC
Extragalactic Database (NED) as well as data from the fifth data
release of the Sloan Digital Sky Survey.  This work has been supported
by the University of Wyoming Research Experience for Undergraduates
program under NSF REU grant AST-0353760.  JCT acknowledges support
from a Zwicky fellowship from the Inst.~of Astronomy, ETH Z\"urich and 
CLAS, University of Florida.

\clearpage
	
\clearpage
	
\clearpage

\clearpage

\clearpage

\clearpage	

\clearpage

\clearpage
	\newcommand{\h}{^h}
\newcommand{\m}{^m}
\newcommand{\dg}{\arcdeg}
\newcommand{\arcm}{\arcmin}

\begin{deluxetable}{cccccccc}
\tabletypesize{\scriptsize}
\tablecolumns{8}
\tablewidth{0pc}
\tablecaption{Properties of the Galaxies Used in this Survey\label{tab1}}
\tablehead{
	\colhead{Galaxy} & \colhead{R.A.} & \colhead{Dec.} & 
	\colhead{Hubble Stage} & \colhead{$M_B$} & \colhead{$v$} & \colhead{Distance} & \colhead{$N_{cluster}$} \\
	 & \multicolumn{2}{c}{(J2000.0)} & \colhead{(Type)} & & \colhead{(km s$^{-1}$)} & 
	\colhead{(Mpc)} & \colhead{M $>10^{5}$ M$_\sun$} \\
	}
\startdata 
\cutinhead{Irregulars}
NGC 2552 & $08\h19\m20\fs53$ & $+50\dg00\arcm34\farcs7$ &  9.0 (SA(s)m)  & -17.3 & 699 & 9.32 & 5 \\
NGC 3738 & $11\h35\m48\fs79$ & $+54\dg31\arcm26\farcs0$ & 10.0 (Im)        & -16.8 & 464 & 6.19 & 8 \\
NGC 4190 & $12\h13\m44\fs73$ & $+36\dg38\arcm02\farcs9$ & 10.0 (Im(p))     & -14.8 & 415 & 5.53 & 7 \\
NGC 4214 & $12\h15\m39\fs17$ & $+36\dg19\arcm36\farcs8$ & 10.0 (IAB(s)m) & -17.1 & 291 & 2.98 & 9 \\
NGC 4449 & $12\h28\m11\fs90$ & $+44\dg05\arcm39\farcs6$ & 10.0 (ImB)       & -18.8 & 422 & 5.63 & 55 \\
NGC 4485 & $12\h30\m31\fs13$ & $+41\dg42\arcm04\farcs2$ & 10.0 (ImB(sp))  & -17.5 & 690 & 9.20 & 90 \\
NGC 4656 & $12\h43\m57\fs73$ & $+32\dg10\arcm05\farcs3$ & 9.0 (SB/Sm(p)) & -18.7 & 646 & 8.61 & 65 \\
IC 3521     & $12\h34\m39\fs50$ & $+07\dg09\arcm37\farcs0$ & 10.0 (ImB)       & -16.0 & 663 & 8.84 & 1 \\
UGC 6541 & $11\h33\m28\fs90$ & $+49\dg14\arcm14\farcs0$ & 10.0 (Im)      & -13.0 & 249 & 3.32  & 3\\ 
UGC 7408 & $12\h21\m15\fs01$ & $+45\dg48\arcm40\farcs8$ & 10.0 (ImA)    & -16.4 & 683 & 9.11 & 2\\
DDO  165 & $13\h06\m24\fs85$ & $+67\dg42\arcm25\farcs0$ & 10.0 (Im)       & -15.3 & 312 & 4.16 & 1 \\
DDO  167 & $13\h13\m22\fs73$ & $+46\dg19\arcm13\farcs2$ & 10.0 (Im)       & -11.6 & 400 & 5.33 & 0 \\
DDO  168 & $13\h14\m27\fs95$ & $+45\dg55\arcm08\farcs9$ &  10.0 (ImB)    & -16.1 & 426 & 5.68 & 1 \\
\cutinhead{Spirals}
NGC 4571 & $12\h36\m56\fs37$ & $+14\dg13\arcm02\farcs5$ & 6.5 (SAd(r))     & -17.0 & 435 & 5.80 & 2 \\
NGC 4713 & $12\h49\m57\fs87$ & $+05\dg18\arcm41\farcs1$ & 7.0 (SABd(rs)) & -17.7 & 715 & 9.54 & 33 \\
NGC 5457 & $14\h03\m12\fs59$ & $+54\dg20\arcm56\farcs7$ & 6.0 (SABcd(rs)) &-20.8 & 503 & 6.71 & 379 \\
\enddata
\tablerefs{Coordinates and radial velocities were taken from NASA/IPAC Extragalactic Database (NED), and morphologies are from RC3.  Distances are from LEDA \citep{LEDA} except for \citet{MCM02} (NGC 4214).  $M_B$s are derived from the apparent B-band magnitudes found in RC3 and our adpoted distances.}
\end{deluxetable}

\clearpage
\begin{deluxetable}{rccccccccc}
\tablecolumns{10}
\tabletypesize{\footnotesize}
\tablewidth{0pc}
\tablecaption{Detected Star Clusters from the Galaxies Found in the SDSS DR5 Data \label{tab2}}
\tablehead{
        \colhead{ID} & \colhead{RA} & \colhead{Dec.} & \colhead{$A_{V}$} &\colhead{$(M_{u'})_o$} & \colhead{$(M_{g'})_o$} & \colhead{$(M_{i'})_o$} & \colhead{$(M_{z'})_o$} & \colhead{Age} & \colhead{Mass} \\
         & \multicolumn{2}{c}{(J2000.0)} & \colhead{(mag)} & & & & & \colhead{(Myr)} & \colhead{$log(M/M_\sun)$} \\
}
\startdata

\cutinhead{NGC 2552 ($A_{V,Gal.}=0.154^1, Z_{fit}=0.020$)}
   3 & 08~19~17.29 & +50~00~14.3 &  0.00 & -11.40$\pm$0.02 & -11.78$\pm$0.01 & -11.04$\pm
$0.02 & -11.20$\pm$0.08 & $ 6.71^{+1.00}_{-1.00}$ & $4.84^{+0.22}_{-0.07}$ \\
   6 & 08~19~18.51 & +50~00~38.7 &  0.00 &  -9.62$\pm$0.05 &  -9.71$\pm$0.03 &  -9.51$\pm
$0.04 & -10.03$\pm$0.15 & $13.01^{+1.40}_{-1.40}$ & $4.56^{+0.11}_{-0.13}$ \\
   7 & 08~19~18.71 & +50~00~40.5 &  0.84 & -11.00$\pm$0.06 & -11.03$\pm$0.03 & -10.10$\pm
$0.03 & -10.71$\pm$0.13 & $ 5.81^{+1.00}_{-1.00}$ & $4.64^{+0.21}_{-0.05}$ \\
   8 & 08~19~20.53 & +50~00~34.7 &  2.52 & -14.43$\pm$0.03 & -14.40$\pm$0.01 & -12.03$\pm
$0.01 & -13.42$\pm$0.03 & $ 4.41^{+1.00}_{-1.00}$ & $5.90^{+0.11}_{-0.15}$ \\
  11 & 08~19~24.48 & +50~00~15.4 &  0.93 & -11.35$\pm$0.04 & -11.18$\pm$0.02 &  -9.85$\pm
$0.03 & -10.03$\pm$0.22 & $ 3.71^{+1.00}_{-1.00}$ & $4.58^{+0.14}_{-0.14}$ \\
  13 & 08~19~20.56 & +50~00~51.3 &  0.01 &  -9.20$\pm$0.08 &  -9.37$\pm$0.05 &  -9.62$\pm
$0.04 &  -9.85$\pm$0.18 & $ 8.41^{+1.60}_{-1.00}$ & $4.27^{+0.12}_{-0.14}$ \\
  16 & 08~19~17.53 & +50~01~27.1 &  0.28 & -10.16$\pm$0.05 & -10.05$\pm$0.04 &  -9.14$\pm
$0.07 &  -8.92$\pm$0.51 & $ 4.01^{+1.10}_{-1.00}$ & $4.14^{+0.15}_{-0.26}$ \\
  17 & 08~19~21.43 & +50~01~38.0 &  4.68 & -15.77$\pm$0.17 & -15.74$\pm$0.02 & -12.06$\pm
$0.01 & -14.57$\pm$0.02 & $ 4.31^{+1.00}_{-1.00}$ & $6.43^{+0.11}_{-0.16}$ \\

\cutinhead{NGC 3738 ($A_{V,Gal.}=0.034^1, Z_{fit}=0.004$)}
   2 & 11~35~45.59 & +54~31~43.8 &  3.46 & -11.25$\pm$0.54 & -10.96$\pm$0.08 &  -8.16$\pm
$0.06 & -10.08$\pm$0.14 & $ 4.71^{+1.50}_{-1.60}$ & $4.67^{+0.22}_{-0.21}$ \\
   5 & 11~35~45.75 & +54~31~38.0 &  2.25 & -12.63$\pm$0.05 & -12.85$\pm$0.02 & -11.31$\pm
$0.02 & -12.81$\pm$0.02 & $10.21^{+1.10}_{-1.10}$ & $5.66^{+0.11}_{-0.07}$ \\
   6 & 11~35~46.78 & +54~31~07.7 &  1.23 &  -9.21$\pm$0.19 &  -9.35$\pm$0.05 &  -8.46$\pm
$0.05 &  -9.49$\pm$0.12 & $10.41^{+13.30}_{-1.10}$ & $4.30^{+0.38}_{-0.10}$ \\
   7 & 11~35~46.51 & +54~31~36.9 &  1.17 & -12.68$\pm$0.02 & -12.46$\pm$0.02 & -11.14$\pm
$0.02 & -11.62$\pm$0.03 & $ 4.91^{+1.00}_{-1.00}$ & $5.26^{+0.06}_{-0.17}$ \\
   9 & 11~35~46.27 & +54~31~52.0 &  1.19 & -10.80$\pm$0.06 & -10.97$\pm$0.03 &  -9.82$\pm
$0.03 & -10.64$\pm$0.06 & $13.01^{+1.40}_{-1.40}$ & $5.02^{+0.11}_{-0.13}$ \\
  10 & 11~35~46.92 & +54~31~41.6 &  1.31 & -12.59$\pm$0.03 & -12.56$\pm$0.02 & -11.15$\pm
$0.02 & -11.56$\pm$0.04 & $ 5.51^{+1.00}_{-1.00}$ & $5.26^{+0.05}_{-0.19}$ \\

\enddata
\tablecomments{$A_{V}$ values of zero indicate that removing additional reddeing only move the cluster further from the Starburst99 model tracks.}
\tablenotetext{1}{The table appears in its entirety in the electron edition.}
\tablenotetext{2}{$A_{V,Gal.}$ from NASA/IPAC Extragalactic Database (NED)}
\end{deluxetable}

\begin{landscape}
\clearpage
	\begin{deluxetable}{crr}
\tabletypesize{\scriptsize}
\tablecolumns{7}
\tablewidth{0pc}
\tablecaption{Rejection Summary \label{tab3}}
\tablehead{
	\colhead{Filter} & \multicolumn{2}{c}{Number of Clusters} \\
	 & \colhead{Irregulars}  & \colhead{Spirals} \\
}
\startdata
Initial sample				& 4,447 & 1,738 \\
Poor photometry				& 3,337 &  664 \\
Rejection by color$^3$			&   23 &   25 \\
Old ($> 20$ Myr)			&  486 &  329 \\
Low mass ($M < 10^{4.4}~M_\sun$)	&  279 &  361 \\
High mass ($M > 10^{7.5}~M_\sun$)	&    1 &    1 \\
Final sample				&  321 &  358 \\
\enddata
\tablecomments{\textit{Poor Photometry} refers to objects rejected due to photometric errors $> 10$\% or having an aperture $>1.5$ times the median stellar aperature.  \textit{Rejection by color$^3$} refers to objects rejected by the color$^3$ routine either by emission line contaimination or lying more than 3$\sigma$ and 0.50 mag from the Starburst99 models.}
\end{deluxetable}

\clearpage
\end{landscape}
	\begin{deluxetable}{cccccccc}
\tablecolumns{8}
\tabletypesize{\footnotesize}
\tablewidth{0pc}
\tablecaption{Completeness Corrections for the Galaxy Samples\label{tab4}}
\tablehead{
        \colhead{Galaxy} & \colhead{Distance} & \multicolumn{6}{c}{Cluster Recovery Rate} \\
 	& \colhead{(Mpc)} & \colhead{$10^{4.40-4.60} M_\sun$} & \colhead{$10^{4.60-4.80} M_\sun$} & \colhead{$10^{4.80-4.99} M_\sun$} & \colhead{$10^{4.99-5.19} M_\sun$} & \colhead{$10^{5.19-5.39} M_\sun$} & \colhead{$10^{5.39-5.58} M_\sun$} \\
	 ~ & ~ & \colhead{$10^{5.58-5.78} M_\sun$} & \colhead{$10^{5.78-5.98} M_\sun$} & \colhead{$10^{5.98-6.18} M_\sun$} & \colhead{$10^{6.18-6.37} M_\sun$} & \colhead{$10^{6.37-6.57} M_\sun$} \\
	}
        \startdata 
	\cutinhead{Irregulars}
	NGC\ 3738 & 6.19  & 0.74 & 0.86 & 0.93 & 0.93 & 0.96 & 0.97 \\
		~ & ~     & 0.97 & 0.96 & 0.97 & 0.97 & 0.96 & \\
	NGC\ 4190 & 5.53  & 0.97 & 0.97 & 0.99 & 1.00 & 0.99 & 1.00 \\
		~ & ~  & 0.98 & 0.98 & 0.99 & 0.99 & 1.00 & \\
	NGC\ 4214 & 2.98  & 0.97 & 0.97 & 0.98 & 0.98 & 0.97 &0.98 \\
		~ & ~  & 0.98 & 0.97 & 0.97 & 0.97 & 0.92 & \\
	NGC\ 4449 & 5.63  & 0.85 & 0.89 & 0.92 & 0.94 & 0.96 & 0.96 \\
		~ & ~  & 0.98 & 0.99 & 0.97 & 0.98 & 0.98 & \\
	NGC\ 4485 & 9.20  & 0.69 & 0.87 & 0.93 & 0.94 & 0.94 & 0.95 \\
		~ & ~  & 0.96 & 0.96 & 0.96 & 0.96 & 0.95 & \\
	NGC\ 4656 & 8.61  & 0.83 & 0.92 & 0.94 & 0.96 & 0.95 & 0.96 \\
		~ & ~  & 0.97 & 0.97 & 0.97 & 0.98 & 0.97 & \\
	UGC\ 6541 & 3.32  & 0.97 & 0.95 & 0.99 & 0.99 & 0.96 & 0.99 \\
		~ & ~  & 0.92 & 0.91 & 0.94 & 0.97 & 0.96 & \\
	DDO\ 168  & 5.68  & 1.00 & 0.99 & 1.00 & 0.99 & 1.00 & 1.00 \\
		~ & ~  & 0.99 & 1.00 & 0.99 & 0.99 & 0.99 & \\
	Average & - & 0.79 & 0.90 & 0.94 & 0.95 & 0.96 & 0.96 \\
		~ & ~  & 0.97 & 0.97 & 0.97 & 0.97 & 0.96 & \\
	\cutinhead{Spirals}
	NGC\ 4571 & 5.80  & 0.99 & 1.00 & 1.00 & 1.00 & 1.00 & 1.00 \\
		~ & ~ & 1.00 & 1.00 & 1.00 & 1.00 & 1.00 & \\
	NGC\ 4713 & 9.54  & 0.67 & 0.87 & 0.97 & 0.98 & 0.99 & 1.00\\
		~ & ~ & 0.99 & 0.99 & 1.00 & 1.00 & 1.00 & \\
	NGC\ 5457 & 6.71  & 0.97 & 0.97 & 0.98 & 0.98 & 0.98 & 0.98 \\
		~ & ~ & 0.98 & 0.98 & 0.98 & 0.98 & 0.98 & \\
	Average & - & 0.92 & 0.96 & 0.98 & 0.98 & 0.99 & 0.99 \\
		~ & ~ & 0.99 & 0.98 & 0.99 & 0.99 & 0.99 & \\
	\enddata
\tablecomments{The average for the corrections is weighted by the number of young ($\la20$~Myr) clusters in each galaxy.}
\end{deluxetable}

\end{document}